\documentclass[aps,prd,10pt,twocolumn,showkeys,superscriptaddress,amsmath,amssymb,nofootinbib]{revtex4-2}

\usepackage{xcolor}
\usepackage{graphicx}
\usepackage{dcolumn}
\usepackage{bm}
\usepackage{tikz}
\usepackage{placeins}
\usepackage{array}
\usepackage{booktabs}
\usepackage{multirow}
\usepackage{hyperref}
\usepackage{orcidlink}
\usepackage{times}

\newcommand{\GRACE}{\texttt{GRACE}}
\newcommand{\FIL}{\texttt{FIL}}

\newcommand{\eg}{e.g., ~}
\newcommand{\ie}{i.e., ~}

\definecolor{purple1}{HTML}{9654D6}

\begin{document}

\title{ \textbf{\texttt{GRACE}: An Open-Source Framework for GPU-Accelerated Numerical Relativity} }

\author{Carlo Musolino\,\orcidlink{0000-0002-9955-3451}}
\email{carlo.musolino@aei.mpg.de}

\affiliation{Max-Planck-Institut f\"{u}r Gravitationsphysik (Albert-Einstein-Institut), Am M\"{u}hlenberg 2, D-14476 Potsdam-Golm, Germany}

\author{Christian Ecker\,\orcidlink{0000-0002-8669-4300}}
\affiliation{Institut f\"{u}r Theoretische Physik, Goethe Universit\"{a}t, Max-von-Laue-Str. 1, 60438 Frankfurt am Main, Germany}

\author{Konrad Topolski\,\orcidlink{0000-0001-9972-7143}}
\affiliation{Institut f\"{u}r Theoretische Physik, Goethe Universit\"{a}t, Max-von-Laue-Str. 1, 60438 Frankfurt am Main, Germany}

\author{Marie Cassing\,\orcidlink{0000-0002-0244-2983}}
\affiliation{Institut f\"{u}r Theoretische Physik, Goethe Universit\"{a}t, Max-von-Laue-Str. 1, 60438 Frankfurt am Main, Germany}

\author{Keneth Miler\,\orcidlink{0009-0005-6955-9894}}
\affiliation{Institut f\"{u}r Theoretische Physik, Goethe Universit\"{a}t, Max-von-Laue-Str. 1, 60438 Frankfurt am Main, Germany}

\author{Harry Ho-Yin Ng\,\orcidlink{0000-0003-3453-7394}}
\affiliation{TAPIR, Mailcode 350-17, California Institute of Technology, Pasadena, CA 91125, USA}

\author{Khalil Pierre} \affiliation{Institut f\"{u}r Theoretische Physik, Goethe Universit\"{a}t, Max-von-Laue-Str. 1, 60438 Frankfurt am Main, Germany}

\author{Elias R. Most\,\orcidlink{0000-0002-0491-1210}}
\affiliation{TAPIR, Mailcode 350-17, California Institute of Technology, Pasadena, CA 91125, USA}

\affiliation{Walter Burke Institute for Theoretical Physics, California Institute of Technology, Pasadena, CA 91125, USA}

\author{Luciano Rezzolla\,\orcidlink{0000-0002-1330-7103}}

\affiliation{Institut f\"{u}r Theoretische Physik, Goethe Universit\"{a}t, Max-von-Laue-Str. 1, 60438 Frankfurt am Main, Germany}

\affiliation{Department of Mathematics, New Uzbekistan University, Tashkent 100007, Uzbekistan}

\affiliation{School of Mathematics, Trinity College, Dublin 2, Ireland}


\date{\today}

\begin{abstract}
We present \texttt{GRACE}, a new GPU-accelerated numerical-relativity framework designed to run efficiently on heterogeneous high-performance computing platforms. Developed from scratch and built exclusively on open-source libraries, \texttt{GRACE} employs \texttt{Kokkos} for performance portability across CPU and GPU architectures and \texttt{p4est} for adaptive mesh refinement. The code evolves the equations of ideal GRMHD---with divergence-free magnetic fields maintained by constrained transport---self-consistently coupled to the Einstein equations in the Z4c formulation, on fixed or adaptively refined grids. We validate the implementation against a suite of standard tests, ranging from magnetized shock tubes and the magnetic rotor in flat spacetime, through (magnetized) Bondi accretion onto a Schwarzschild black hole and the ringdown of a perturbed spinning puncture, to neutron-star oscillation spectra in fixed and dynamical spacetimes and the merger of binary black holes. As more demanding applications, we evolve two binary neutron-star mergers---an equal-mass, unmagnetized system with an ideal-gas equation of state and an unequal-mass, magnetized system with a finite-temperature tabulated equation of state---finding the inspiral dynamics to agree well with the \texttt{FIL}~code. We also report single-device throughput together with strong- and weak-scaling results on multiple GPU and CPU architectures. \texttt{GRACE} is publicly released together with \texttt{GRACEpy}, a basic post-processing and data-analysis environment.
\end{abstract}

\keywords{numerical relativity, general relativistic magnetohydrodynamics, high-performance computing, neutron stars, gravitational waves}

\maketitle

\section{Introduction}
\label{sec:intro}

Seminal advances in the early 1960s and 1970s on the dynamics of general relativity~\cite{Arnowitt:1962hi} laid the mathematical groundwork for the field of numerical relativity (NR), which has since grown and evolved through countless theoretical and technical breakthroughs. Among the many milestones, a few are particularly significant: the first simulations of binary neutron star (BNS) mergers in the late 1990s~\cite{Shibata:1999wm}; the first successful simulations of black-hole inspiral and merger in the mid-2000s~\cite{Pretorius:2005gq, Alcubierre:2004hr, Campanelli:2005dd, Baker_2006}; and the first multiphysics simulations of BNS mergers incorporating magnetic fields and weak interactions~\cite{Sekiguchi:2011zd}.

Arguably, the field has now reached a level of maturity at which it can deliver not only qualitative insights but also quantitative predictions that connect theoretical models of relativistic astrophysical systems to observations. Modern applications of NR are numerous and diverse, ranging from the production of accurate waveform models for compact binary inspirals~\cite{Dietrich:2018phi, Kiuchi_2020, Foucart:2020xkt, Scheel:2025jct}, to the modeling of BNS mergers (see reviews~\cite{Baiotti:2016qnr, Radice:2020ddv}) with complex microphysics and magnetic fields~\cite{Kiuchi:2015qua, Kiuchi_2025, Most:2018eaw, Bauswein:2018bma, Ecker:2019xrw, Weih:2019xvw, Bauswein:2020ggy, Blacker:2020nlq, Foucart:2020qjb, Most:2021zvc, Tootle:2022pvd, Kiuchi:2023obe, Ecker:2024kzs, Aguilera-Miret:2025nts},
to the simulation of disks surrounding supermassive black holes~\cite{Fragile:2007dk, Porth:2020txf, Fragile:2025nes,Wong:2025ily, Zhang:2025uug}, and many more.

Progress in this field has often been driven not only by innovations in numerical methods but also by the rapid growth of the amount of resources available through high-performance computing (HPC) facilities. The most recent of these advancements has been the advent of heterogeneous computing platforms over the course of the last decade. These consist of systems where one or more graphics processing units (GPUs) supplement the computational power of the traditional CPU-based system, acting as specialized accelerators which can perform massively parallel computations very efficiently.
 
An increasing number of NR codes have already been modified to run efficiently on GPUs (see, e.g.,~\cite{Liska:2019uqw, Shankar:2022ful, Stone2024, Kalinani:2024rbk, Zhu:2024utz, Palenzuela:2025ucx,han2026sacrakperformanceportablenumericalrelativity}), demonstrating that this technology has the potential of enabling longer, more accurate and more realistic simulations of various astrophysical phenomena~\cite{Lalakos:2025msz, Wang:2025yco, Gutierrez:2026ngt, Radice:2025djo}. Nevertheless, adopting GPU accelerated computing in NR presents several challenges, since many of the commonly employed techniques and algorithms well adapted to CPUs perform poorly on GPU architectures.

This paper is aimed at presenting a novel NR simulation framework designed to meet these challenges. \GRACE\ (General Relativistic Astrophysics Code for Exascale) was built from the ground up with efficiency and portability on modern HPC systems in mind. It achieves this by leveraging the \texttt{Kokkos}~library~\cite{Trott2021, Trott2022_etal}, a minimal abstraction layer that exposes common parallel loop paradigms and convenient memory allocation functionality.

The numerical grid in \GRACE\ is constructed following a forest-of-octree strategy, which allows for locally refining and coarsening the grid with minimal algorithmic overhead. The basic octree bookkeeping algorithms are executed on CPU using the \texttt{p4est}~open source library~\cite{Burstedde2011, carsten_burstedde_2024_10839051}, whereas all data operations are done on GPU using custom-written kernels. \GRACE\ supports both fixed (FMR) as well as adaptive mesh refinement (AMR).

In its current form, \GRACE\ is written to solve the equations of ideal general-relativistic magnetohydrodynamics (GRMHD), \ie for a plasma with infinite conductivity, either in fully backreacted, dynamical spacetimes or, alternatively, within the Cowling approximation, in which the spacetime metric is kept fixed as a prescribed background.

This paper is organized as follows: Section~\ref{sec:equations} provides an overview of the Z4c formulation of the Einstein equations and the basic equations of GRMHD. Section~\ref{sec:methods} gives a brief description of the relevant algorithms and numerical methods, as well as the adaptations that were necessary to efficiently deploy them on GPU systems. Section~\ref{sec:code} describes the structure and implementation of \GRACE, including the numerical grid, adaptive mesh refinement and the task-based ghost-zone update. Section~\ref{sec:tests} presents the results of a series of tests aimed at validating the implementation of~\texttt{GRACE}~involving both isolated and compact binary objects, including a comparison between results obtained with \GRACE\ and those from the \texttt{Frankfurt/IllinoisGRMHD}~(\texttt{FIL})~GRMHD code~\cite{Most:2019kfe}. Section~\ref{sec:perf} demonstrates the performance and scaling of \GRACE, and we close with conclusions in Sec.~\ref{sec:conclusions}.

Throughout this manuscript, Greek spacetime indices run from $0$ to $3$, with the time component denoted interchangeably by $0$ and $t$, while Latin spatial indices run from $1$ to $3$. We adopt geometrized units with $c = G = 1$. In the tests involving matter at astrophysical scales---i.e.\ the neutron-star and binary-neutron-star simulations---we additionally set $M_\odot = 1$ and quote masses in solar masses; the flat-spacetime and vacuum tests are scale-invariant, and there we work in code units with the mass scale fixed as stated in each case.

\section{Mathematical Setup}
\label{sec:equations}

We will now briefly review the basic equations of the Z4c system and ideal GRMHD implemented in \GRACE, and refer readers interested in a more detailed discussion to specialized review articles and textbooks~\cite{Font:2000pp, Anton:2005gi, Gourgoulhon:2007ue, BonaPalenzuelaBonaCasas2009, Rezzolla:2013dea, Lehner:2014asa, Baiotti:2016qnr, Shibata2016NumericalRelativity, Mewes:2020vic, Baumgarte:2021skc, Bambi:2025btf}.


\subsection{Einstein equations in a 3+1 decomposition}
In the following, we assume that spacetime with metric $g_{\mu\nu}$ is foliated by a family of spacelike hypersurfaces. Denoting by $n^\mu$ the future-directed unit normal to these hypersurfaces, we write it in terms of the lapse $\alpha$ and shift $\beta^i$ as
\begin{equation}
  n^\mu = \left( 1/\alpha, -\beta^i / \alpha \right) \,.
\end{equation}
The induced spatial metric $\gamma_{\alpha\beta}$ is then obtained as the projector onto the subspace orthogonal to $n^\mu$,
\begin{equation}
  \gamma_{\alpha\beta} := g_{\alpha\beta} + n_\alpha n_\beta \,.
\end{equation}

In \GRACE, the spacetime evolution is achieved by solving the Z4 formulation of the Einstein equations~\cite{Bona:2003fj, Alic:2011a} and, in particular, using the Z4c system~\cite{Hilditch:2012fp}. Following~\cite{Shibata2016NumericalRelativity}, we adopt $\widetilde{W}$ as the conformal factor, so that the physical metric is expressed in terms of the conformal metric $\tilde{\gamma}_{ij}$ as
\begin{equation}
\gamma_{ij} := \widetilde{W}^{-2} \tilde{\gamma}_{ij} \,,
\end{equation}
where the latter has unit determinant, \ie $\tilde{\gamma}=1$. The conformal traceless extrinsic curvature is then given by
\begin{equation}
\tilde{A}_{ij} := \widetilde{W}^{2} \, ( K_{ij} - \frac{1}{3} K \,
\gamma_{ij} ) \,,
\end{equation}
where $K :=\gamma^{ij} K_{ij}$ is the trace of the extrinsic curvature tensor. We recall that following usual conventions, the indices of conformal tensors (which we indicate with a tilde) are lowered and raised with the conformal metric $\tilde{\gamma}_{ij}$ and its inverse.

We then follow~\cite{Hilditch:2012fp} and introduce the auxiliary degrees of freedom $\Theta$---whose dynamics are prescribed to propagate and damp constraint violations---and $\tilde{\Gamma}^i$, which is initialized according to $\tilde{\Gamma}^i = -\partial_j \tilde{\gamma}^{ij}$. The evolved extrinsic curvature trace is then taken to be $\hat{K} := K - 2\Theta$. Denoting with $\mathcal{D}_i$ the covariant derivative operator compatible with the physical metric $\gamma_{ij}$, and introducing the trace of the conformal Christoffel symbols of the second kind $(\tilde{\Gamma}_d)^i := \tilde{\gamma}^{jk} {\tilde{\Gamma}^i}_{jk}$, which corresponds with the dynamical variable $\tilde{\Gamma}^i$ only on shell, we can write the Einstein equations in the following form
\begin{widetext}
\begin{align}
    \partial_t \widetilde{W} &= \frac{1}{3} \widetilde{W} \left[ \alpha (\hat{K}
      + 2\Theta) - \partial_i \beta^i \right] + \beta^i \partial_i
    \widetilde{W} \,, \label{eq:z4c_w} \\ \partial_t \tilde{\gamma}_{ij} &=
    - 2 \alpha \tilde{A}_{ij} + \beta^k \partial_k \tilde{\gamma}_{ij} +
    2 \tilde{\gamma}_{k(i}\partial_{j)}\beta^k - \frac{2}{3}
    \tilde{\gamma}_{ij} \partial_k \beta^k \,,\label{eq:z4c_gammat}
    \\ \partial_t \hat{K} &= - \mathcal{D}^i \mathcal{D}_i \alpha +
    \alpha \left[ \tilde{A}_{ij} \tilde{A}^{ij} + \frac{1}{3} \left(
      \hat{K} + 2\Theta \right)^2 \right] + 4 \pi \alpha \left( S^{^{\rm ADM}}
      + \rho_{_{\rm ADM}} \right) + \alpha \, \kappa_1 \,(1-\kappa_2)
    \, \Theta + \beta^i \partial_i \hat{K} \,,\label{eq:z4c_Khat}
    \\ \partial_t \tilde{\Gamma}^i &= -2 \tilde{A}^{ij}\partial_j \alpha
    + 2 \alpha \left[ \tilde{\Gamma}^i_{jk} \tilde{A}^{jk} -3
      \tilde{A}^{ij} \partial_j \log{\widetilde{W}} - \frac{1}{3}
      \tilde{\gamma}^{ij} \partial_j \left( 2 \hat{K} + \Theta \right) -
      8\pi \tilde{\gamma}^{ij} S^{^{\rm ADM}}_j \right] \notag \\ &+
    \tilde{\gamma}^{jk} \partial_j \partial_k \beta^i + \frac{1}{3}
    \tilde{\gamma}^{ij} \partial_j \partial_k \beta^k + \beta^j
    \partial_j \tilde{\Gamma}^i - (\tilde{\Gamma}_d)^j \partial_j \beta^i
    + \frac{2}{3} (\tilde{\Gamma}_d)^i \partial_j \beta^j - 2 \alpha \,
    \kappa_1 \left[ \tilde{\Gamma}^i - (\tilde{\Gamma}_d)^i \right] \,
    , \label{eq:z4c_Gamma} \\ \partial_t \Theta &= \frac{1}{2} \alpha
    \left[ R - \tilde{A}_{ij} \tilde{A}^{ij} + \frac{2}{3} \left( \hat{K}
      + 2 \Theta \right)^2 \right] - \alpha \left[ 8\pi \rho_{_{\rm ADM}} +
      \kappa_1 (2 + \kappa_2) \Theta \right] + \beta^i \partial_i \Theta
    \,, \label{eq:z4c_Theta} \\ \partial_t \tilde{A}_{ij} &= \widetilde{W}^2
    \left[\alpha (R_{ij} - 8 \pi S^{^{\rm ADM}}_{ij} ) -\mathcal{D}_i
      \mathcal{D}_j \alpha \right]^{\rm TF} + \alpha \left[ \left(
      \hat{K} + 2\Theta \right) \tilde{A}_{ij} - 2 {\tilde{A}^k}_i
      \tilde{A}_{kj} \right] + \beta^k \partial_k \tilde{A}_{ij} + 2
    \tilde{A}_{k(i}\partial_{j)}\beta^k -\frac{2}{3} \tilde{A}_{ij}
    \partial_k \beta^k \, \label{eq:z4c_At} .
\end{align}
\end{widetext}

In Eqs.~\eqref{eq:z4c_w}--\eqref{eq:z4c_At}, $R_{ij}$ is the Ricci tensor and $R$ its trace, whereas the constants $\kappa_1$ and $\kappa_2$ determine the constraint damping behavior of the Z4c system and are taken to be $\kappa_1=0.02$ and $\kappa_2=0$ unless specified otherwise.

Moreover, we introduced the matter source terms
\begin{align}
\rho_{_{\rm ADM}} &:= n^\mu n^\nu T_{\mu\nu} \,, \label{eq:z4c_rhoadm}
\\ S^{^{\rm ADM}}_i &:= - \gamma^\mu_i n^\nu T_{\mu \nu} \,
, \label{eq:z4c_Si} \\ S^{^{\rm ADM}}_{ij} &:= \gamma^\mu_i \gamma^\nu_j
T_{\mu\nu} \,, \label{z4c_Sij}
\end{align}
and defined $S^{^{\rm ADM}}$ as the trace of $S^{^{\rm ADM}}_{ij}$.

The system of evolved equations is subject to the algebraic constraints ${\rm det}(\tilde{\gamma}_{ij}) = 1$ and $\tilde{\gamma}^{ij} \tilde{A}_{ij} = 0$, as well as the Hamiltonian and momentum constraints
\begin{align}
\mathcal{H} &= R - \tilde{A}^{ij} \tilde{A}_{ij} + \frac{2}{3} K^2 - 16\,
\pi \rho_{_{\rm ADM}} = 0 \,, \label{eq:z4c_hamiltonian} \\ \mathcal{M}^i
&= \partial_j \tilde{A}^{ij} + \tilde{\Gamma}^i_{jk} \tilde{A}^{jk} -
\frac{2}{3} \tilde{\gamma}^{ij} \partial_j \left( \hat{K} + 2\Theta
\right)\nonumber\\ &- 3 \tilde{A}^{ij} \partial_j \log{\widetilde{W}} - 8 \pi
\tilde{\gamma}^{ij} S^{^{\rm ADM}}_j = 0\,. \label{eq:z4c_momentum}
\end{align}

To close the system we need to prescribe a gauge that determines the values of the lapse function $\alpha$ and the shift vector $\beta^i$. We follow the widely adopted ``puncture gauge'' where the lapse is evolved according to the $1+\log$ slicing condition~\cite{Bona:1998dp}
\begin{equation} \label{eq:oplog}
\partial_t \alpha = -2\alpha \hat{K} + \beta^i \partial_i \alpha \,,
\end{equation}
and the shift follows the Gamma-driver equations~\cite{Alcubierre:2002kk, vanMeter:2006vi, Bruegmann:2006ulg}
\begin{align} 
\partial_t \beta^i &= \frac{3}{4} \tilde{B}^i + \beta^k \partial_k
\beta^i \,, \label{eq:gdriver1} \\ \partial_t \tilde{B}^i &= \partial_t
\tilde{\Gamma}^i + \beta^k ( \partial_k \tilde{B}^i - \partial_k
\tilde{\Gamma}^i ) - \eta \tilde{B}^i \,,
\label{eq:gdriver2}
\end{align}
where $\tilde B^i$ is an additional auxiliary degree of freedom which is initialized as zero, and $\eta$ is a free parameter that damps the shift evolution which we typically choose to be $2/M$, where $M$ is the Arnowitt--Deser--Misner (ADM) mass of the spacetime.

\subsection{The equations of GRMHD}

As already mentioned, the equations of GRMHD describe the evolution of a magnetized fluid on a curved background in the limit of infinite conductivity. They can be cast as a set of conservation equations for the fluid energy-momentum tensor and for the conserved particle number (e.g.,~baryon number in nuclear matter), supplemented by the induction equation for the magnetic field. By defining the fluid four-velocity $u^\mu$, as well as its projection onto the spatial hypersurfaces (i.e.,~the three velocity measured by an Eulerian observer)
\begin{equation}\label{eq:three_velocity}
v^i = \frac{1}{\alpha} \left( \frac{u^i}{u^0} + \beta^i \right) \,,
\end{equation}
we can then define the Lorentz factor as $W := (1 - v^i v_i)^{-1/2}$ and we introduce the coordinate velocity $\tilde{v}^i := u^i/u^0$ for future convenience. Under the assumption of a perfect fluid (i.e., neglecting dissipative effects and heat conduction), the energy momentum tensor can be written as
\begin{equation} \label{eq:tmunu}
  T^{\alpha \beta} = \left( \rho h + b^2 \right) u^\alpha u^\beta +
  \left( p + \frac{1}{2} b^2 \right) \, g^{\alpha \beta} - b^\alpha
  b^\beta \,,
\end{equation}
where $u^\mu$ is the fluid four-velocity, $\rho$ is the rest-mass density of the fluid, $p$ the pressure, $h := 1 + \epsilon + p/\rho $ the specific enthalpy (with $\epsilon$ denoting the specific internal energy) and $b^\mu$ the magnetic field in the comoving frame of the fluid, which is related to the Eulerian frame field $B^i$ by
\begin{equation} \label{eq:smallb}
  b^0 = \frac{B^i u_i}{\alpha} \,,\quad b^i = \frac{B^i + \alpha \,b^0\,
    u^i}{W}\,.
\end{equation}
For future convenience, we report here also the square norm of the comoving field
\begin{equation}\label{eq:smallb2}
 b^2 = \frac{B^i B_i + \alpha^2 (b^0)^2}{W^2} \,.
\end{equation}

The conserved rest-mass, energy and momentum densities are defined as
\begin{align}
  D & := W \rho \,,\\
  \tau & := W^2 \left( \rho h + b^2 \right) - \left( p +
  \frac{b^2}{2} \right) - \alpha^2 (b^0)^2 - D \,, \\
  S_i & := W \left(\rho h + b^2 \right) u_i - \alpha b^0 \, b_i \,,
\end{align}
where we note that the rest-mass energy has been subtracted from the total energy density following~\cite{Font:2000pp}.

With these definitions, we can write the equations of ideal GRMHD in their conservative form~\cite{Rezzolla:2013dea}
\begin{equation} \label{eq:grmhd_cons}
  \partial_t ( \sqrt{\gamma} \, \boldsymbol{U} ) + \partial_i ( \sqrt{\gamma}
  \, \boldsymbol{F}^i(\boldsymbol{U}) ) = \sqrt{\gamma} \, (
  \boldsymbol{S}(\boldsymbol{U}) - \boldsymbol{R} ) \,,
\end{equation}
where $\gamma$ is the determinant of the spatial three-metric $\gamma_{ij}$, $\boldsymbol{U}$ is the vector of conserved variables, $\boldsymbol{F}^i$ the corresponding fluxes, and $\boldsymbol{S}$ is the geometric source term. Explicit expressions for these quantities are given by
\begin{widetext}
\begin{equation} \label{eq:grmhd_U_F_S}
\begin{gathered}
\boldsymbol{U} = \begin{bmatrix} D \\ S_j \\ \tau \\ B^j \\
\end{bmatrix}\,,\quad 
\boldsymbol{F}^i = \begin{bmatrix} D \tilde{v}^i \\ \alpha \left\{ \left(
  \rho h + b^2 \right) u^i u_j - b^i b_j + \left( p + {b^2}/{2} \right)
  \delta^i_j \right\} \\ W^2 \left( \rho h + b^2 \right) \tilde{v}^i +
  \left( p + {b^2}/{2} \right) \beta^i - \alpha^2 b^0 b^i - D \tilde{v}^i
  \\ B^j \tilde{v}^i - B^i \tilde{v}^j
\end{bmatrix} \,, \\ 
\boldsymbol{S} = \begin{bmatrix} 0\\ ({\alpha}/{2}) T^{\mu \nu}
  \partial_j g_{\mu \nu} \\ \alpha \left( T^{00} \beta^i \beta^j + T^{0
    i}\beta^j + T^{0 j}\beta^i + T^{ij} \right) K_{ij} - \alpha \left(
  T^{0 i} + T^{00} \beta^i \right) \partial_i \alpha \\ 0
\end{bmatrix} \,.
\end{gathered}
\end{equation}
\vspace{-\baselineskip}
\end{widetext}
The radiation reaction source terms $\boldsymbol{R}$ are set to zero throughout this work.

In addition to the main GRMHD system, \GRACE\ evolves two passive scalars advected by the matter flow, namely the conserved electron fraction $D\,Y_e$ and the conserved specific entropy $D\,s$. Both satisfy a conservation equation of the form
\begin{equation}\label{eq:passive_scalars}
\partial_t \left(\sqrt{\gamma}\, D\,\psi \right) + \partial_i
\left(\sqrt{\gamma}\, D\,\psi\,\tilde v^i \right) =
\sqrt{\gamma}\,\mathcal{R}_\psi \,,
\end{equation}
for $\psi \in \{Y_e,\, s\}$. The source terms $\mathcal{R}_\psi$ are set to zero in the present work, so that the electron fraction $Y_e$ and the specific entropy $s$ are passively transported with the fluid.

The system of equations is closed by providing an equation of state (EOS) relating the pressure to other thermodynamic variables. \GRACE\ supports four families of EOS that cover the range of thermodynamic models needed across the test problems of Sec.~\ref{sec:tests}, as well as the targeted production applications.

The simplest family of EOS is given by an ideal gas with an adiabatic index $\Gamma$, $p = (\Gamma-1)\,\rho\,\epsilon$. For applications in which only the cold part of the EOS is known---such as the construction of equilibrium neutron-star models---the code provides a piecewise polytropic EOS that interpolates between a set of $N$ segments of the form $p_{{\rm cold},i} = K_i \rho^{\Gamma_i}$, with $\rho_{i-1}\leq\rho\leq\rho_{i}$ and $i \in [1, N]$, joined continuously across density breakpoints $\rho_i$~\cite{Read:2008iy}. For each segment $i$, the corresponding specific internal energy of the cold component follows from the first law of thermodynamics as
\begin{equation}
  \epsilon_{{\rm cold},i}(\rho) =\epsilon_{{\rm cold},i}(\rho_{i-1})+\int_{\rho_{i-1}}^{\rho}\frac{p_{{\rm cold},i}(\rho')}{\rho'^2}
  \,d \rho'\,.
\end{equation}

Shock-heating can be approximately captured via a so-called ``hybrid EOS'', that is, by augmenting any cold EOS with a $\Gamma_{\rm th}$-law thermal component,
\begin{equation}\label{eq:gthth_law}
    p_{\rm th} = (\Gamma_{\rm th} - 1)\,\rho\,\epsilon_{\rm th}\,,
\end{equation}
where $\epsilon_{{\rm th}} = \epsilon - \epsilon_{{\rm cold}}$ denotes the thermal part of the specific internal energy and the total pressure is $p = p_{{\rm cold}} + p_{{\rm th}}$. 
For diagnostic purposes, a physical temperature can be associated with the thermal component through the ideal-gas relation
\begin{equation}\label{eq:hybrid_temperature}
k_B\,T = \bar{m}\,(\Gamma_{\rm th}-1)\,\epsilon_{\rm th}\,,
\end{equation}
where $\bar m$ is a dimensionful mass scale that fixes the temperature units. We should emphasize that with the hybrid EOS the evolution depends only on $\epsilon_{\rm th}$ and not on $T$, so that the choice of $\bar m$ is purely a bookkeeping convention used when reporting temperatures in physical units---typically the baryon mass for nuclear-matter applications such as binary neutron-star mergers. The cold EOS in the hybrid model can be either a piecewise polytropic fit or the zero-temperature, beta-equilibrium slice of a fully tabulated EOS.

For applications requiring the full nuclear-physics thermodynamics, \GRACE\ implements a reader for three-dimensional tabulated EOSs in which all relevant thermodynamic quantities are tabulated as functions of rest-mass density, temperature and charge fraction. The currently supported file format is the standardized CompOSE format~\cite{Typel:2013rza, Oertel:2016bki, CompOSECoreTeam:2022ddl}; extensions to other table layouts are straightforward.

Most of the test cases in Sec.~\ref{sec:tests} use the simple polytropic or ideal-gas EOS, with the exception of the magnetized binary neutron star merger, for which a tabulated version of the fully temperature-dependent SFHo model~\cite{Steiner:2012rk} from the CompOSE database is employed.

\subsection{Extraction of Gravitational Waves}
\label{sec:math_GWs}

For gravitational-wave (GW) analysis, the Newman--Penrose scalar $\Psi_4$ is constructed by contracting the Weyl tensor $C_{\alpha\beta\gamma\delta}$ with a complex null tetrad $\{\ell^\mu, m^\mu, \hat{n}^\mu, \bar{m}^\mu\}$ adapted to the spacetime foliation
\begin{equation}
\Psi_4 := C_{\alpha\beta\gamma\delta}\, \hat{n}^\alpha \bar m^\beta \hat{n}^\gamma \bar
m^\delta\,,
\end{equation}
where, recalling that $n^\mu$ denotes the future-directed unit normal to the hypersurface and denoting by $\{u^i, v^i, w^i\}$ an orthonormal spatial triad obtained via Gram--Schmidt orthonormalization of the coordinate radial and angular directions with respect to the physical metric $\gamma_{ij}$, the tetrad legs are constructed as $\ell^\mu = (n^\mu + u^\mu)/\sqrt{2}$, $\hat{n}^\mu = (n^\mu - u^\mu)/\sqrt{2}$, and $m^\mu = (v^\mu + i\,w^\mu)/\sqrt{2}$, with $\bar{m}^\mu$ the complex conjugate of $m^\mu$~\cite{Bruegmann:2006ulg}. With these choices $\ell^\mu$ and $\hat{n}^\mu$ are null and satisfy $\ell^\mu \hat{n}_\mu = -1$, while $m^\mu$ is purely spatial (see, e.g.,~\cite{Baker:2001sf, Bishop2016} for details). Using the symmetries of the Weyl tensor together with standard curvature identities, $\Psi_4$ can be written entirely in terms of fields defined on the spatial hypersurface
\begin{align}
\Psi_4 & := \left( {R}_{ijkl}+2K_{i[k}K_{l]j}\right)
    \hat{n}^i\bar{m}^j\hat{n}^k\bar{m}^l\\
    & \phantom{:=} -8 \left( K_{j[k,l]}+{\Gamma
    }_{j[k}^pK_{l]p}\right) \hat{n}^{[0}\bar{m}^{j]}\hat{n}^k\bar{m}^l\\
    & \phantom{:=} +4\left({R}_{jl}-K_{jp}K_l^p+KK_{jl}\right)
    \hat{n}^{[0}\bar{m}^{j]}\hat{n}^{[0}\bar{m}^{l]}\,,
\end{align}
where $\hat{n}^0 = 1/\sqrt{2}$ and $\hat{n}^i = -u^i/\sqrt{2}$ are the components of $\hat{n}^\mu$ in the frame adapted to the unit normal $n^\mu$, and where indices surrounded by square brackets are anti-symmetrized.

In our implementation, we decompose $\Psi_4$ into complex modes $\psi_4^{\ell m}$ using spin-weighted spherical harmonics ${}_{s}Y_{\ell m}$ with weight $s=-2$, evaluated on a spherical surface centered at the origin of the numerical grid
\begin{equation}
\Psi_4=\sum_{\ell=2}^{\infty}\sum_{m=-\ell}^{m=\ell}\psi_4^{\ell
  m}{_{-2}Y}_{\ell m}\,.
\end{equation}
The plus ($+$) and cross ($\times$) polarizations of the GW strain modes are then obtained from the real and imaginary parts of the twice time-integrated $\psi_4^{\ell m}$ modes
\begin{equation}
h_+^{\ell m}(t) - i h_\times^{\ell m}(t) = \int_{-\infty}^{t} dt'
\int_{-\infty}^{t'} dt'' \, \psi_4^{\ell m}(t'') \,.
\end{equation}

Finally, we define the instantaneous GW phase $\phi^{\ell m}$ and frequency $f^{\ell m}_{\rm GW}=\omega^{\ell m}/(2\pi)$ as follows
\begin{align}
  &\phi^{\ell m} := \arctan \frac{h_\times^{\ell m}}{h_+^{\ell m}}\,,\\
  &\omega^{\ell m} := \dot{\phi}^{\ell m}:=\frac{d\phi^{\ell m}}{dt}\,.
\end{align}

\section{Numerical Methods}
\label{sec:methods}

In what follows we describe briefly the numerical methods for the solution of the combined set of the Z4c and GRMHD equations, highlighting also the handling of some important aspects of the numerical integration, such as the use of a constrained-transport scheme, the conversion of the conservative variables to the primitive ones, and the use of suitable boundary conditions.

\subsection{Numerical solution of the Einstein equations}
\label{ssec:z4c_discretization}

The Z4c equations are solved by central finite-differencing with upwind stencils for the shift-advection terms. \texttt{GRACE} supports both $4{\rm th}$- and $6{\rm th}$-order accurate discretizations of the Z4c right-hand-side, with the latter being the default choice for the simulations presented in this work unless otherwise noted. We also add a Kreiss--Oliger dissipation operator~\cite{KreissOliger1973GARP} of the form
\begin{equation}\label{eq:ko_diss}
\partial_t u \rightarrow \partial_t u + \frac{(-1)^{r+1}\,\epsilon_{\rm
    diss}\, h^{2r-1}}{2^{2r}} \sum_i D^{(2r)}_i u\,,
\end{equation}
where $h$ denotes the grid-spacing, $D^{(2r)}_i$ is a centered finite difference approximation of the $2r$-th partial derivative along direction $i$, $\epsilon_{\rm diss}$ is a constant dissipation amplitude and $u$ is any of the evolved Z4c variables. Following~\cite{Husa:2007hp}, we choose $r=4$ for the $6{\rm th}$-order discretization and $r=3$ for the $4{\rm th}$-order one, which guarantees that the dissipation term does not spoil the formal convergence-order of the underlying finite-difference scheme. Furthermore, the algebraic constraints of unit determinant of the conformal metric and trace-free condition of $\tilde{A}_{ij}$ are imposed after each sub-step of the time integrator, as well as after the interpolation of data following the modification of the grid or the filling of ghost-zones.

\subsection{Numerical solution of the matter equations}
\label{ssec:riemann}

The MHD solver in \GRACE\ is based on standard high-resolution shock-capturing (HRSC) methods using an approximate Riemann solver on a Cartesian grid (see, \eg \cite{Font:2000pp} for a pedagogical introduction). Adopting a conservative formulation of the GRMHD equations, the numerical flux at each interface needed for the Riemann problem is computed from the reconstructed left/right primitive states using either a Harten--Lax--van Leer--Einfeldt (HLLE)~\cite{Harten83,Einfeldt88} approximate Riemann solver or a local Lax--Friedrichs (LLF, also known as the Rusanov flux~\cite{RUSANOV1962304}) flux. For a conserved variable $U$ with physical flux $F$ in the direction $i$, the HLLE flux reads
\begin{equation}\label{eq:hlle_flux}
    F^{\rm HLLE} = \frac{c_{\max} F_L + c_{\min} F_R - c_{\max}
      c_{\min}\,(U_R - U_L)}{c_{\max} + c_{\min}},
\end{equation}
where the bounding speeds are constructed from the characteristic speeds $\lambda_\pm$ on either side of the interface as
\begin{equation}\label{eq:hlle_speeds}
    c_{\max} = \max\!\left(0,\, \lambda_+^L,\, \lambda_+^R\right), \qquad
    c_{\min} = -\min\!\left(0,\, \lambda_-^L,\, \lambda_-^R\right),
\end{equation}
so that $c_{\max}, c_{\min} \ge 0$ by construction. The clipping against zero ensures that the flux reduces to the correct upwind value when the flow is supersonic in either direction. The LLF flux is recovered by replacing both bounding speeds with the single largest characteristic speed at the interface, $c_{\max} = c_{\min} = \max(|\lambda_\pm^L|, |\lambda_\pm^R|)$.

The characteristic speeds $\lambda_\pm$ are the fast magnetosonic speeds of the GRMHD system in the direction $i$~\cite{Gammie:2003rj}. Defining the Alfv\'en speed and the combined fast speed
\begin{equation}\label{eq:fast_speed}
    v_{\rm A}^2 = \frac{b^2}{b^2 + \rho h}, \qquad v_0^2 = v_{\rm A}^2 +
    c_s^2\,(1 - v_{\rm A}^2).
\end{equation}
The speeds $\lambda_\pm$ are the roots of the quadratic equation
\begin{equation}\label{eq:wavespeed_quadratic}
    \mathcal{A}\, \lambda^2 + \mathcal{B}\, \lambda + \mathcal{C} = 0,
\end{equation}
with coefficients
\begin{align}
    \mathcal{A} & := (u^0)^2\,(1 - v_0^2) + \frac{v_0^2}{\alpha^2},
    \\ \mathcal{B} & := 2\left( \frac{\beta^i}{\alpha^2}\, v_0^2 -
    (u^0)^2\, \tilde{v}^i\, (1 - v_0^2) \right), \\ \mathcal{C} & :=
    (u^0)^2\, (\tilde{v}^i)^2\, (1 - v_0^2) - v_0^2 \left[ \gamma^{ii} -
      \left( \frac{\beta^i}{\alpha} \right)^2 \right].
\end{align}
Here, $\tilde{v}^i$ is the advection velocity in the grid frame introduced in Sec.~\ref{sec:equations}, $u^0 = W/\alpha$ the time component of the fluid four-velocity, and $\gamma^{ii}$ is the diagonal component of the inverse spatial metric in the direction $i$. The larger root defines $\lambda_+$ and the smaller $\lambda_-$, which enter Eq.~\eqref{eq:hlle_speeds} directly.

A subset of the primitive variables (see Sec.~\ref{sec:c2p} for details) is reconstructed from cell centers to cell interfaces using one of a hierarchy of operators: a piecewise-constant (``donor cell'') reconstructor, a slope-limited piecewise-linear scheme with either the minmod or the monotonized-central limiter~\cite{VANLEER1977263}, and the third- and fifth-order weighted essentially non-oscillatory schemes of Jiang \& Shu~\cite{JIANG1996202} supplemented with the WENO-Z smoothness indicators of~\cite{BORGES20083191, Acker2016}. Unless stated otherwise, the simulations presented in this work employ the fifth-order WENO-Z reconstruction. Concretely, the reconstructed quantities at each interface are the rest-mass density $\rho$, the velocity vector $z^i$, the electron fraction $Y_e$, the specific entropy $s$, one of the two thermodynamic primitives $T$ or $p$, and the two transverse components of the magnetic field $B^i$; the remaining thermodynamic primitives are recovered locally from the EOS at the reconstructed states. Unless otherwise stated, all tests perform the reconstruction on the pressure.

The component of $B^i$ normal to the interface is not reconstructed but instead read directly from the face-staggered storage, a requirement of the constrained-transport scheme that guarantees a unique normal magnetic field at every cell face and is essential for preserving the discrete divergence-free constraint to round-off (see Sec.~\ref{sec:CT} for details). Furthermore, to safeguard the scheme against the appearance of unphysical states in regions of strong gradients, low density or near floors, \GRACE\ implements an \textit{a posteriori} first-order flux correction (FOFC) of the kind introduced for \texttt{AthenaK} by~\cite{Fields:2024pob}.

The algorithm proceeds in two passes per Runge-Kutta (RK) substage (see also Sec.~\ref{sec:GTI}). In the first pass, the high-order fluxes are computed at every cell interface, including their contributions to the face-staggered electromotive forces (EMF) of the constrained-transport sector, and a tentative in-place update of both the conserved fluid variables and the staggered magnetic field is performed without the geometric source terms.

The conservative-to-primitive inversion (see also Sec.~\ref{sec:c2p}) is then attempted on this tentative state, and every cell in which the inversion triggers the atmosphere floor or otherwise returns a nonphysical primitive set is flagged. In the second pass, the high-order fluxes at the interfaces surrounding each flagged cell are replaced by a first-order accurate flux, and the EMFs at every edge that touches a flagged cell are recomputed consistently from the corrected face fluxes; the constrained-transport update is then redone at those edges before the geometric sources are added and the final update is applied. The first-order scheme consists of an LLF flux computed on donor-cell reconstructed data. The local fallback to a first-order scheme is therefore restricted to the cells in which the higher-order solution would have been unphysical, while full WENO-Z accuracy is preserved everywhere else. We note that, when employing tabulated equations of state, the temperature falling below the atmosphere floor does not trigger the first-order correction. This is done since we have found that in problems such as binary neutron-star mergers these failures can sometimes trigger in the interior of stars during the inspiral, where employing a first-order accurate scheme might significantly degrade the quality of the solution. We also implement the relaxed discrete maximum principle (DMP) limiter of~\cite{Fields:2024pob} (see also~\cite{Zanotti_2015}, where a cell is flagged for FOFC if the conserved density $D$ or energy $E:=\tau+D$ exceed the maximum (or fall below the minimum) of all its $26$ neighbors at the previous time-step by a prescribed amount. We chose to apply this principle to $E$ as opposed to $\tau$ since the latter can be negative.

Additionally, \GRACE\ supports the option of adding even more dissipation at near-atmosphere densities, where $|\lambda|$ can be small (for stationary unmagnetized flows it reduces to the sound speed of the fluid), by replacing the wave-speed in the Rusanov flux with the speed of light in the grid frame $\lambda_c = |\beta^i| + \alpha\, \sqrt{\gamma^{ii}}$.

\subsection{Constrained transport}
\label{sec:CT}

To preserve the divergence-free condition of the magnetic field, the ideal GRMHD solver in \texttt{GRACE} is based on the constrained transport (CT) approach, where the densitized magnetic field $\sqrt{\gamma} \, B^i$ is evolved at the cell faces, and where the electric field used in the induction equation, which in ideal MHD is just
\begin{equation}
E_i = -\sqrt{\gamma} \, \epsilon_{ijk} \tilde{v}^j B^k \,,
\end{equation}
is defined at cell edges. This ensures that the divergence of the magnetic field, defined as
\begin{equation}
\nabla \cdot \boldsymbol{B} = \frac{\partial_i (\sqrt{\gamma} B^i
  )}{\sqrt{\gamma}} \,,
\end{equation}
remains constant within round-off error. Naturally, the electric field itself must be unique at each cell edge and upwind with respect to the governing equations.

\texttt{GRACE} supports two different methods of computing the electric field: the so-called ``upwind constrained transport'' (UCT) of~\cite{Londrillo:2003qi, DelZanna:2007pk, Mignone2021}---in which the coordinate velocity and the fastest wave-speeds are stored at cell faces during the fluid update, reconstructed to the edges together with the staggered magnetic field, and combined there via a four-state generalization of the HLL Riemann solver---and the method of~\cite{Gardiner_2008} (see also~\cite{Kiuchi_2022}), which we will refer to as CT-contact following~\cite{Mignone2021}. Since all the tests of Sec.~\ref{sec:tests} employ the CT-contact scheme unless stated otherwise, we limit the detailed description to the latter and refer the reader to the references above for the UCT algorithm.

In the CT-contact scheme the electric field is computed at cell faces using the Riemann solver, and the final value of the electromotive force is obtained by averaging the four values coming from the faces that share that edge and correcting with derivative terms that are upwinded with respect to the contact speed of the adjacent faces. In the formulas that follow we adopt the convention that $E^z$ with three integer indices denotes the cell-centered electric field, while a half-integer index in a given coordinate direction indicates the value of $E^z$ stored at the corresponding cell face. As an example, for the $z$ component of the electric field this formula reads
\begin{align}
  \label{eq:ct_contact}
  &E^z_{i-1/2,j-1/2,k} = \nonumber \\
  &\frac{1}{4} \Big( E^z_{i-1/2,j,k} + E^z_{i-1/2,j-1,k} +
  E^z_{i,j-1/2,k} + E^z_{i-1,j-1/2,k} \Big) \nonumber \\
  &+ \frac{1}{8} \left( \delta_y
  E^z\big|_{i-1/2,j-3/4,k} - \delta_y E^z\big|_{i-1/2,j-1/4,k} \right)
  \nonumber \\
  &+ \frac{1}{8} \left( \delta_x E^z\big|_{i-3/4,j-1/2,k} - \delta_x
  E^z\big|_{i-1/4,j-1/2,k} \right) \,.
\end{align}
Here, the upwind-selected $y$-difference operator at the north position is defined as
\begin{align}\label{eq:e_diff}
  & \delta_y E^z\big|_{i-1/2,j-1/4,k} := \nonumber \\
  & \phantom{+\,\,} \left(1-s^x_{i-1/2,j,k} \right)
  \left( E^z_{i,j,k} - E^z_{i,j-1/2,k} \right) \nonumber \\
  & + \left(1+s^x_{i-1/2,j,k} \right) \left( E^z_{i-1,j,k} -
  E^z_{i-1,j-1/2,k} \right) \,,
\end{align}
with $s^x_{i-1/2,j,k}$ denoting the sign of the conserved mass flux at the $x$-face $(i-1/2,j,k)$. Each of the two terms in Eq.~\eqref{eq:e_diff} is the $y$-difference between the cell-centered $E^z$ at $(i,j,k)$ or $(i-1,j,k)$ and the face-centered $E^z$ at the corresponding $y$-face $(i,j-1/2,k)$ or $(i-1,j-1/2,k)$; the convex combination of indicator functions $(1\mp s^x)$ selects the upstream cell relative to the contact wave at the $x$-face that the edge belongs to. The companion expression for $\delta_y E^z|_{i-1/2,j-3/4,k}$ at the south position is obtained by shifting the cell-centered $y$-index from $j$ to $j-1$ and updating both the cell-centered and face-centered $E^z$ values consistently (with the upwinding evaluated at the corresponding south $x$-face $(i-1/2,j-1,k)$), and the two $\delta_x E^z$ operators in Eq.~\eqref{eq:ct_contact} follow by an analogous $x\leftrightarrow y$ exchange.

We note that the CT-contact scheme requires the storage of the electric field at cell centers for the evaluation of Eq.~\eqref{eq:e_diff}, but does not require the storage of wave-speeds at cell faces, so that the two schemes have an essentially equivalent memory footprint.

\subsection{Conservative-to-primitive inversion}
\label{sec:c2p}

As is the case for the vast majority of GRMHD solvers, the equations are solved for the ``conserved'' variables defined in Eq.~\eqref{eq:grmhd_cons}, but the evaluation of the right-hand side of those equations requires the knowledge of so-called ``primitive'' variables such as the rest-mass density $\rho$, the fluid velocity $v^i$ and so on. The conversion of the conserved variables into primitive variables is normally referred to as the ``conservative-to-primitive'' (C2P) procedure. Since the definition of conserved variables involves the metric and the EOS, its inverse is not algebraic and generally requires the solution of a nonlinear root-finding problem. Numerous methods exist in the literature to solve this problem~\cite{Siegel_2018, Galeazzi_2013} and \GRACE\ follows the one of~\cite{Kastaun:2020uxr}, which was shown to be very robust in regimes where the problem is ill conditioned such as high Lorentz factors or large magnetizations~\cite{Cheong:2020kpv, Ng_2024}. Furthermore, the method reduces to the solution of a one-dimensional root-finding problem, which can be efficiently tackled even in GPU codes where instruction-level divergence is an important performance concern.

In particular, the set of primitive variables stored at cell centers in \GRACE\ consists of the rest-mass density $\rho$, the pressure $p$, the specific internal energy $\epsilon$, the temperature $T$, the electron (or charge) fraction $Y_e$, the vector $z^i=W\, v^i$, and the cell-centered magnetic field $B^i$ (obtained from the face-staggered representation by simple area averaging). This set of primitives is recovered after each RK sub-step and the root-finding is performed by means of a bracketed Brent iteration.

Whenever this procedure fails---for example because the conserved variables fall outside the physical domain of the EOS, or the bracket becomes degenerate---the code optionally falls back on the entropy-based inversion of~\cite{Gammie:2003rj}, in which the conserved entropy replaces $\tau$ in the recovery and the resulting purely thermodynamic problem is solved by a one-dimensional bracketed iteration. In particular, the entropy based fallback uses the same master function as the main inversion, but replaces the energy equation with the entropy. The latter being trivially recoverable from the conserved state, this procedure avoids having to recover the specific internal energy from $\tau$, instead computing the pressure directly from the tentative rest-mass density and entropy per baryon. The inverter calls the EOS through a uniform set of accessors, so that the same recovery routine is used for any of the EOS families introduced in Sec.~\ref{sec:equations}.

If all the inversion procedures fail, the cell is set to an ``atmosphere'' condition, where the rest-mass density and temperature are fixed to (small) constant values $\rho_{\rm atmo}$ and $T_{\rm atmo}$, the velocity in the Eulerian frame is set to zero, the electron fraction (if present in the EOS) is set to its beta-equilibrium value, and the rest of the primitives are consistently reconstructed from the EOS.

Before attempting the inversion, conservative variables are checked for unphysical states and potentially corrected. In particular, if $D<\sqrt{\gamma} \, \max(0, \rho_{\rm atmo})$ the cell is set to atmosphere without attempting the inversion. The conserved energy is bounded from below by a floor value, $\tau \geq \tau_{\rm fl}$, with
\begin{equation}
    \tau_{\rm fl} = \frac{1}{2} B^i B_i + \min(0, \epsilon_0) \, D \,,
\end{equation}
where $\epsilon_0$ is the smallest specific internal energy allowed by the EOS at the given $Y_e$ and at $\rho = D/\sqrt{\gamma}$ (which is an upper bound on $\rho$, since $W\geq 1$). Finally, the conserved momentum is limited to the region allowed by the dominant energy condition~\cite{Galeazzi_2013}
\begin{equation}
    S_i S^i \leq \left( \tau + D \right)^2 \,.
\end{equation}

After the inversion is successful and a set of primitives has been obtained, the cell is set to atmosphere if $\rho \leq \rho_{\rm atmo} ( 1 + \delta_{\rm atmo} )$, where $\delta_{\rm atmo}$ is a configurable parameter which we typically set to $0.1$, and all the conserved variables are overwritten to be consistent with this state. If instead the recovered temperature is below the atmosphere value $T < T_{\rm atmo}$, only $T,\epsilon$ and the pressure are overwritten, while the density and velocity are left unchanged. In this case, only $\tau$ and $S_i$ are adjusted to be consistent with the primitive state. At this stage we also check if the cell should be excised, which can be decided based on coordinates $r \leq r_{\rm excision}$ or based on the lapse $\alpha\leq \alpha_{\rm excision}$, in which case the primitive and conservative states are floored in a similar way to what is done for atmosphere cells. The code supports the option of relaxing these floors in the intermediate steps of the RK time-stepper, which we employ in the binary neutron-star merger test with tabulated EOS. This is done because we have found that when cells are marginally above the atmosphere density and surrounded by atmosphere, the state being floored in one or more of the sub-stages caused a nonconservative injection of energy which led to spurious heating of marginally above threshold cells.

Finally, to avoid numerical instabilities, we limit the maximum Lorentz factor $W\leq W_{\rm max}$ (typically set to $W_{\rm max}=50$) and rescale the velocity if this bound is exceeded. In magnetized flows, we also limit the magnetization $\sigma = b^2/\rho$ to be below a certain configurable maximum, which we normally take to be $\sigma_{\rm max} = 100$ and we increase the density slightly when this bound is exceeded. Since these limiting procedures modify the rest-mass density and the Lorentz factor, all conserved quantities need to be adjusted to remain consistent.

\subsection{General time integration}
\label{sec:GTI}

The semi-discrete evolution equations are integrated in time with an explicit RK scheme. The available choices are a forward Euler step, a midpoint (RK2) scheme, the third-order strong-stability-preserving scheme (SSPRK3) of Shu \& Osher~\cite{SHU1988439}, and the four-stage fourth-order low-storage RK method of Ketcheson~\cite{KETCHESON20101763} in its three-register ``3S*'' form, which we adopt as the default RK4 integrator. Compared to the classical fourth-order RK scheme, the 3S* formulation reduces the number of full state-vector copies that need to be held in device (CPU or GPU) memory during a step from five to three, which is a non-negligible benefit on GPUs where the evolved state of the coupled Z4c--GRMHD system is the largest single memory user. Implicit-explicit (IMEX) variants are also implemented for problems that include radiation transport, but are not exercised in the present work. The time-step size is determined globally from a Courant--Friedrichs--Lewy (CFL) condition based on the fastest characteristic speed across the grid; subcycling in time across refinement levels is not employed.

\subsection{Conservation across refinement interfaces}
\label{ssec:refluxing}

To ensure that the conservative nature of the GRMHD equations is preserved across mesh refinement boundaries, we follow the strategy commonly adopted in adaptive finite-volume codes (see e.g.,~\cite{BERGER198964}) and recompute the flux densities on the coarse side of fine--coarse interfaces using the fine-grid fluxes. The corrected coarse flux at the matching face is taken to be the area-weighted average of the four fine-grid fluxes overlapping it, and replaces the original coarse flux \textit{in place} before the divergence is evaluated. This in-place replacement makes the procedure bit-invariant under repartitioning of the underlying grid via a message-passing interface (MPI).

An analogous re-circulation is applied to the upwind EMFs computed by the CT scheme: the EMFs on the four short fine edges that share a coarse edge are summed, divided by the number of fine intervals, and the result is used to replace the coarse EMF before the magnetic field is advanced. Together with the CT discretization, this ensures that the discrete divergence-free magnetic-field constraint $\nabla\cdot\boldsymbol{B}=0$ is preserved across refinement boundaries at the level of round-off error.

\subsection{Outer boundary conditions}
\label{ssec:outer_bcs}

For Z4c variables the outer boundaries of the computational domain are filled with a Sommerfeld-type radiative boundary condition,
\begin{equation}\label{eq:sommerfeld}
\partial_t u = -v\,s^i \partial_i u + \frac{v}{r}\,(u_\infty -
u) \,,
\end{equation}
where $s^i$ is the unit radial vector in coordinate space, $u_\infty$ is the asymptotic value of the variable and $v$ is the characteristic speed of the outgoing wave.

We adopt $v=1$ for the conformal metric components $\tilde\gamma_{ij}$, the conformal traceless extrinsic curvature $\tilde A_{ij}$, the conformal connection $\tilde\Gamma^i$ and the constraint-damping variable $\Theta$, which all carry the physical GW content of the system. For the lapse $\alpha$ and the trace $\hat K$, however, the principal symbol of the system under the $1+\log$ slicing condition~\cite{Bona:1998dp, Hilditch:2012fp} couples the two variables so that they propagate as gauge modes with characteristic speed $\sqrt{2}$. The Sommerfeld kernel is therefore applied to these two variables with $v=\sqrt{2}$, which is the proper outgoing-wave speed for the gauge sector and avoids partial reflection of the gauge front back into the computational domain. The shift sector $\beta^i$, $\tilde B^i$ is treated in the same way at $v=1$, since the Gamma-driver does not introduce a nontrivial characteristic speed at the linearised level.

Spatial derivatives at the boundary are evaluated at second-order accuracy with centered or one-sided stencils depending on the cell index. For the hydrodynamic variables, \GRACE\ provides zeroth-order (donor-cell) and third-order extrapolation boundary kernels, with the former used as the default outflow boundary condition. Reflection symmetry across coordinate planes can additionally be enforced for variables of any tensor rank, and is employed in several of the tests of Sec.~\ref{sec:tests} to reduce the computational cost.

\section{Code Structure and Implementation}
\label{sec:code}

All numerical work on-device, \ie CPU or GPU, in \GRACE\ is expressed through the \texttt{Kokkos}~\cite{Trott2021, Trott2022_etal} parallel-loop and memory abstractions. The same source tree thus targets multiple accelerator and CPU backends; in practice, the results presented in this work were obtained with the CUDA, HIP and OpenMP backends.

\subsection{Numerical spatial grid}

The numerical grid in \GRACE\ is constructed as a forest of octrees managed through the \texttt{p4est} library~\cite{Burstedde2011}, which provides an efficient distributed-memory implementation of the underlying connectivity, neighbor look-ups, tree-balancing and load-balancing partition operations. \texttt{p4est} is run exclusively on the host, while all data-bearing operations on grid cells happen on device.

Each individual leaf of the tree, which we refer to as a \textit{quadrant}, consists of an equal number $n$ of cells per spatial dimension, supplemented by two sets of ghost-zones on either side. These ghost-zones are not evolved themselves, but represent an overlap of each quadrant with its neighbors, thus allowing for the interpolation of data and evaluation of differential operators which require stencils.

The base grid in \GRACE\ can either consist of one or more octrees at a uniform refinement level $\ell$, in which case each tree consists of $8^\ell$ quadrants, or of a nonuniform grid specified by a series of fixed mesh refinement boxes. The grid is always $2:1$ and is balanced for all neighbors, that is, if two quadrants share a face, edge or corner, then their refinement level can differ at most by one.

\subsection{Adaptive mesh refinement}

During the evolution, a refinement criterion can be specified to periodically decide which quadrants to refine or coarsen. This criterion is usually taken to be an error estimate based on derivatives of evolved quantities, the value of a tracking field such as the conformal factor, or the proximity of compact objects identified by their center of mass; the code is easily extended to custom criteria. Once quadrants have been flagged, the modification of the forest itself---refinement and coarsening, restoration of the 2:1 balance, and repartitioning across MPI ranks for load balance via \texttt{p4est}'s space-filling-curve ordering---is performed by \texttt{p4est} code on the host; \GRACE\ then fills the data of the newly created quadrants on device, as described below.

In the case of refinement one quadrant---the parent---is replaced by eight children which are one level finer, i.e.,~have half the cell spacing. The data from the parent quadrant is then used to fill the newly created cells in a process commonly referred to as ``prolongation'', and during coarsening the newly created quadrant's data is obtained through the inverse operation, ``restriction''. In \GRACE,\ both operations happen on device via custom-written kernels, and the order of the interpolation can be specified independently for each evolved variable, reflecting the different smoothness properties of the underlying fields.

For variables evolved by the finite-volume hydrodynamic scheme, the code provides a second-order slope-limited prolongation and a conservative cell-volume--weighted restriction, both of which preserve the cell-averaged values exactly. For the Z4c metric variables, which are evolved with finite differences, both prolongation and restriction employ centered five-point Lagrange interpolation, i.e.,~a degree-four polynomial reconstruction whose truncation error is of fifth order in the grid spacing (see also~\cite{Stone_2020, Zhang:2025uug}). For the staggered magnetic field, special care has to be taken whenever any form of interpolation is performed to ensure that no violations of the divergence-free constraint are introduced.

In the case of restriction this is fairly straightforward, with the coarse magnetic field value being obtained as the surface-weighted average of the values at the fine cell faces. Prolongation, on the other hand, is significantly more involved, and we follow the divergence-preserving prescription of Ref.~\cite{TOTH2002736} (see also~\cite{Kiuchi_2012,Stone_2020}), which constructs the fine staggered fields by combining a second-order centered reconstruction of the coarse field with a discrete curl correction that exactly cancels any monopole contribution. This prolongation operator is paired with the EMF re-circulation procedure of Sec.~\ref{ssec:refluxing}, so that the discrete divergence-free constraint is propagated across both static and dynamic refinement boundary changes.

Generally speaking, the magnetic field at evolved faces (including the ones at the interface of two quadrants, which overlap) is never modified in-place, since this would lead to a violation of the divergence-free constraint. The only exception to this rule is applied when a new quadrant of level $\ell$ is created during mesh refinement, and one of its face-neighbors was already at level $\ell$. Indeed, naively applying the prolongation formula at the face where these two quadrants touch would preserve the cell-wise divergence-free property of the magnetic field, but does not guarantee that the net magnetic flux across the face is zero. Instead, in this case we simply copy the data from the already-fine neighbor.

\subsection{Ghost-zone updates}

After each update of the interior cells, ghost-zones must be updated to allow for a self-consistent calculation of all quantities to the next timelevel. This process is fairly involved and we will limit ourselves to a short overview of the necessary steps and how they are implemented in \GRACE. Additional and more detailed information can be found in the code documentation.

Firstly, ghost-zones can be separated into interior and physical boundaries. The former are simply an overlap of each quadrant with its neighbor, which can either be at the same refinement level, one level higher, or one level lower. The latter represent cells that sit outside the grid, and whose values must be prescribed according to the boundary conditions described in Sec.~\ref{ssec:outer_bcs}. Interior ghost-zones where the neighbor quadrant is at the same refinement level can simply be copied if both quadrants belong to the same MPI rank, or exchanged via asynchronous MPI operations if this is not the case.

When two adjacent quadrants are at different levels, the ghost-zone filling procedure largely follows the one presented in Ref.~\cite{Stone_2020}. In such case, interior data of the fine quadrants is restricted into so-called ``coarse buffers'', each of which consists of a smaller quadrant with $n/2$ cells plus ghost-zones in each direction. At this point, the coarse neighbor can simply fill its own ghost-zones using data from the coarse buffers, and fill the coarse buffers' ghost-zones using its own interior data. Since prolongation of fine ghost-zones requires a stencil of valid data, this operation needs to wait until all ghost-zones of the coarse buffer are filled via copy or application of outer boundary conditions.

It is important to remark that an efficient implementation of the ghost-zone filling procedure is crucial for achieving good performance for hyperbolic PDE solvers, in particular when the main computation happens on GPU where the arithmetic throughput is very high. For this reason, in \GRACE\ we adopt a custom task-based execution model for this section of the code, where each independent kernel represents a task and all tasks are collected in a ``directed acyclic graph'' (DAG), whose execution fills all the ghost-zones.

The DAG is rebuilt only when the grid topology changes (i.e.,~after a regrid) and its execution carries no per-step bookkeeping overhead. To construct it, a per-quadrant neighbor table is assembled, leveraging \texttt{p4est}'s efficient top-down search algorithms and topology-aware iteration infrastructure~\cite{IsaacBursteddeWilcoxEtAl15}. After all the neighbor data is collected, the DAG can be constructed by tracing all dependencies that need to be fulfilled before the data across a specific quadrant's faces, edges and corners can be retrieved. Communications are split on a per-rank basis, where all the data going to and coming from a given remote MPI rank is placed in a single contiguous buffer, and data dependencies are reflected in task dependencies. For example, all data which needs to be copied across local quadrant faces is collected in a single task, which has no dependencies. Similarly, data that needs to be copied across faces belonging to remote MPI ranks are grouped in one task per rank, so that when each individual message is successfully delivered the data can immediately be used by its dependencies.

Furthermore, tasks are grouped according to whether they are exclusively memory-bound or whether they utilize floating-point units. For example, copy tasks only perform a handful of integer operations to compute cell indices and are dominated by the cost of loading and storing data from global memory, whereas prolongation of the magnetic field components involves a non-negligible amount of floating-point work. These different groups of tasks are then asynchronously scheduled on different execution streams on device, so that their execution can overlap. \GRACE\ maintains a certain number of streams active at all times in a pool, so that work from different groups can be scheduled on these persistent streams in a round-robin fashion.

\subsection{Initial data}
\label{ssec:initial_data}

\GRACE\ provides native readers for the most commonly used open-source initial-data solvers in numerical relativity. For binary neutron-star and mixed binary configurations the code interfaces directly with the \texttt{FUKA} library~\cite{Papenfort2021b, Tootle2021}, which is built on \texttt{KADATH}~\cite{Grandclement:2009ju} and represents the spectral solution on a multi-domain compactified grid. \GRACE\ ingests the resulting fields by evaluating the spectral expansion at the desired Cartesian coordinates. Alternative quasi-equilibrium binary configurations can be loaded from solutions produced with \texttt{LORENE}~\cite{Gourgoulhon:2000nn} through an analogous interface.

\begin{figure*}[htb]
\centering \includegraphics[width=\textwidth]{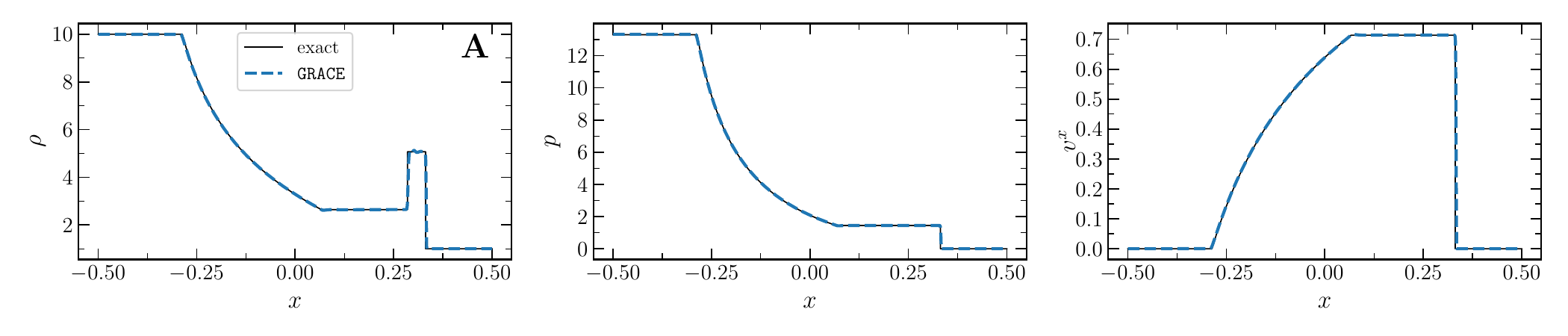}
\includegraphics[width=\textwidth]{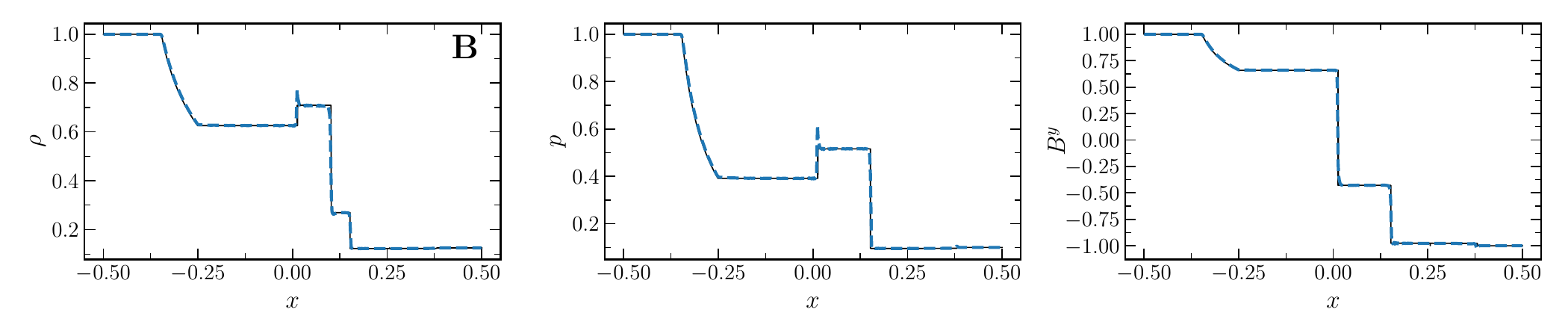}
\includegraphics[width=\textwidth]{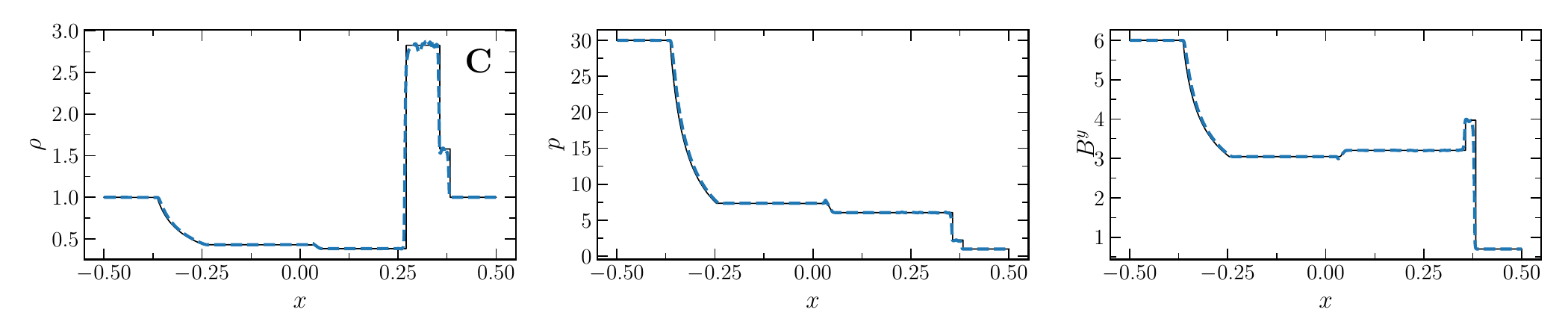}
\caption{Top to bottom: solutions of the shock-tube problems A, B, and C from Tab.~\ref{tab:shocktube_id} at $t=0.4$. In each panel the thin solid black line is the exact solution and the blue dashed line the \GRACE\ result; the exact solutions are those of Refs.~\cite{Rezzolla_2003} (setup~A) and~\cite{Giacomazzo_2006} (setups~B and~C).}
\label{fig:shocktube}
\end{figure*}

Binaries in vacuum are instead imported from the \texttt{TwoPunctures}~\cite{Ansorg:2004ds} pseudospectral solver, statically linked into \GRACE\ and called at initialization time on the host. Single-star and accretion-disk configurations such as the TOV models and Bondi setups of Sec.~\ref{sec:tests} are computed on the fly inside the code itself.

Finally, \GRACE\ exposes a runtime configuration interface based on the YAML\footnote{YAML Ain't Markup Language, a human-readable data serialization standard; see \url{https://yaml.org}.} format that controls all simulation parameters, writes diagnostic and analysis output in HDF5 format, and provides a checkpoint/restart mechanism that allows long simulations to be resumed without loss of accuracy.

\section{Code Tests}
\label{sec:tests}

In what follows, we will present a comprehensive series of validation tests for the implementation of the \texttt{GRACE} code, beginning with a series of tests of the MHD solver on flat spacetime, followed by evolutions of isolated and binary compact objects, both in vacuum and with matter content.

\subsection{Shock-tubes}

As a first test of the magnetohydrodynamics solver in \GRACE, we present a series of shock-tube solutions. The initial left and right states are listed in Table~\ref{tab:shocktube_id}.
\begin{table}[htb]
\begin{tabular}{ccccccccccc}
\toprule setup & $\Gamma$ & side & $\rho$ & $p$ & $B^x$ & $B^y$ & $B^z$
\\ \midrule \multirow{2}{*}{A} & \multirow{2}{*}{$5/3$} & L & $10.0$ &
$13.33$ & $0.0$ & $0.0$ & $0.0$ \\ & & R & $1.0$ & $6.6\times 10^{-7}$ &
$0.0$ & $0.0$ & $0.0$ \\ \midrule \multirow{2}{*}{B} &
\multirow{2}{*}{$2$} & L & $1.0$ & $1$ & $0.5$ & $1.0$ & $0.0$ \\ & & R &
$0.125$ & $0.1$ & $0.5$ & $-1.0$ & $0.0$ \\ \midrule \multirow{2}{*}{C} &
\multirow{2}{*}{$5/3$} & L & $1.0$ & $30.0$ & $5.0$ & $6.0$ & $6.0$ \\ &
& R & $1.0$ & $1.0$ & $5.0$ & $0.7$ & $0.7$ \\ \bottomrule
\end{tabular}
\caption{Initial conditions (in code units) for the shock-tube tests presented in the text and shown in Fig.~\ref{fig:shocktube}. The left (L) configurations correspond to the initial states for $x\leq 0$ while the right (R) ones to those at $x>0$. The initial velocities are set to zero for all setups. The second column from the left reports the adiabatic index used for the ideal-gas EOS.}\label{tab:shocktube_id}
\end{table}
The corresponding numerical results, together with the closed-form solutions, are shown in Fig.~\ref{fig:shocktube}.

\begin{figure}[h!tb]
\centering \includegraphics[width=0.48\textwidth]{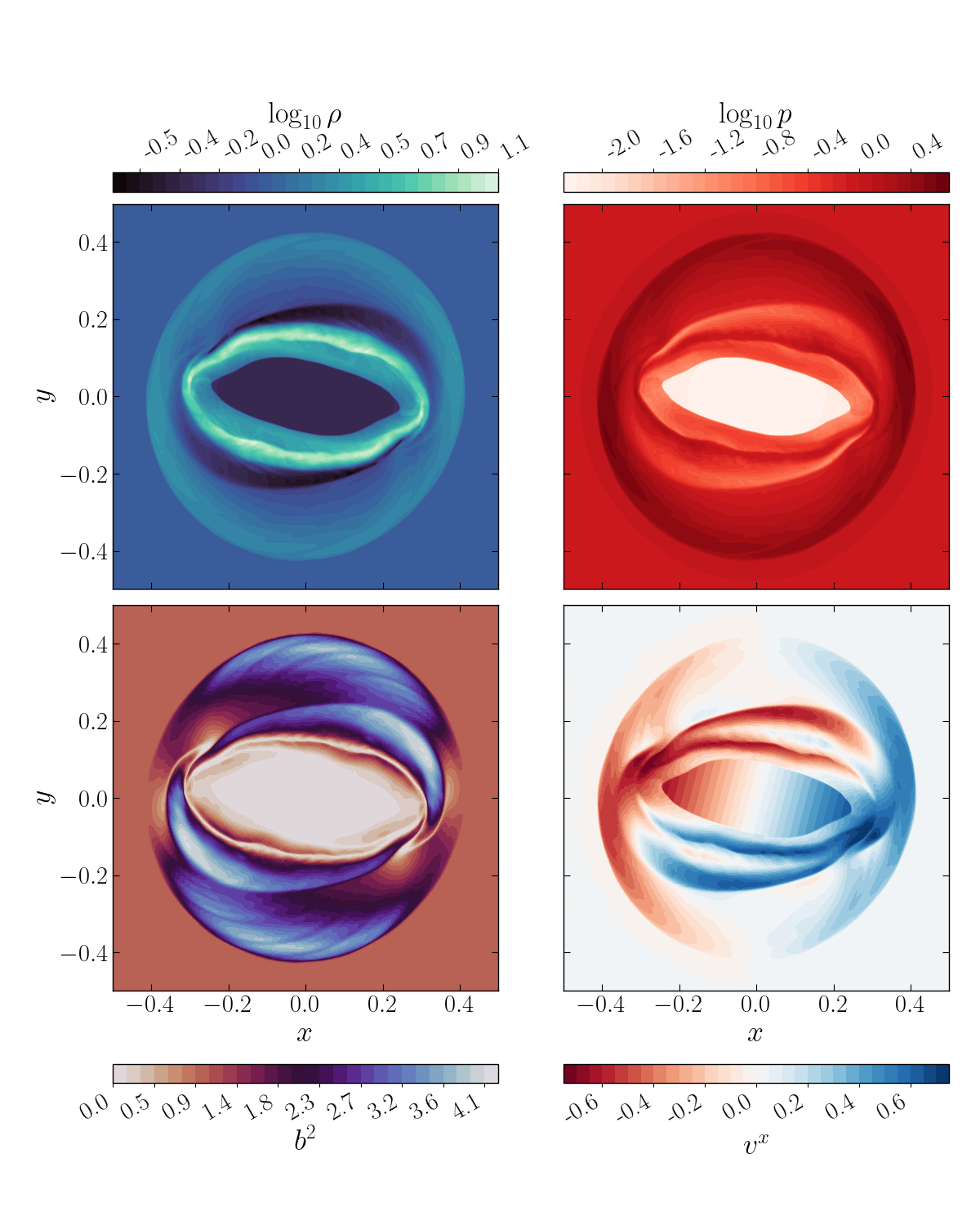}
\caption{Numerical solution for the magnetic rotor test at time $t=0.4$ obtained with \GRACE: rest-mass density (upper left), pressure (upper right), comoving magnetic-field strength squared (lower left), and $x$-velocity of the fluid (lower right).}
\label{fig:rotor}
\end{figure}

In particular, setup A consists of a pure hydrodynamical shock-tube resulting in a right-moving blast wave and a left-moving rarefaction wave. Setups B and C, instead correspond to test cases 1 and 2 from Refs.~\cite{Balsara2001, Giacomazzo:2005jy, Giacomazzo:2007ti} and both involve nonzero magnetic fields. The exact solutions are obtained with the exact Riemann solvers of Refs.~\cite{Rezzolla_2003, Giacomazzo_2006}. All tests are discretized on a grid spanning $x\in [-0.5,0.5]$ covered by $1280$ cells and solved with a third-order RK time integrator and a fixed CFL factor of $0.25$. The solution for setup A at $t=0.4$ is shown in the top row of Fig.~\ref{fig:shocktube}, where the code can be seen to accurately capture the moderately relativistic blast wave propagating to the right in the computational domain, as well as the rarefaction wave moving to the left from the initial discontinuity.

The density, pressure and $y$-component of the magnetic field in shock tube B, also at $t=0.4$, are shown in the middle row of Fig.~\ref{fig:shocktube}. The situation is analogous for setup C, shown in the bottom row of Fig.~\ref{fig:shocktube}, where the solution is again well captured despite small amplitude oscillations appearing in the density around the contact discontinuities. We attribute these spurious oscillations to the high-order WENO-Z reconstruction employed. Overall, the behavior reported in Fig.~\ref{fig:shocktube} provides evidence that \GRACE\ is able to correctly reproduce discontinuous flows both in relativistic hydrodynamics and in relativistic MHD.

\subsection{Magnetic rotor}

The next test we consider is the magnetized-rotor problem first introduced in Ref.~\cite{1999JCoPh.149..270B} and later extended to relativistic MHD by Ref.~\cite{Del_Zanna_2003} (see also~\cite{Neuweiler:2024jae}). In this setup, a dense, uniformly rotating cylinder is embedded in a nonrotating, lower-density ambient medium. More specifically, the rotating core is confined within a cylindrical radius $r = 0.1$, with density $\rho_{\mathrm{in}} = 10$ and angular velocity $\Omega = 9.95$, corresponding to a maximum fluid velocity of $v_{\max} = 0.995$. The surrounding medium has a rest-mass density $\rho_{\mathrm{out}} = 1$. An ideal-gas EOS with $\Gamma = 5/3$ is employed, together with a uniform initial pressure $p = 1$. Both the cylinder and the ambient medium are threaded by an initially uniform magnetic field, with components $B^x = 1$ and $B^y = B^z = 0$. The system is then evolved until $t = 0.4$. Torsional Alfv\'en waves are expected to develop, thus removing angular momentum from the rotor and increasing the magnetic pressure near its edge due to the winding of the magnetic field. This increased pressure is then expected to compress the rotor itself, deforming it.

The solution obtained by \GRACE\ on a numerical grid spanning $(x,y)\in[-0.5,0.5]$ with a uniform spacing $h=1/512\simeq 0.002$ is shown in Fig.~\ref{fig:rotor}. We note that the problem is symmetric under translations in the $z$ direction, we therefore set up a minimal set of points along this direction and impose periodic boundary conditions. The solution was once again obtained with a third-order RK scheme and a CFL factor of $0.25$. The advantage of this test is that it allows us to evaluate the ability of the code in a highly nonlinear MHD regime. The disadvantage, however, is that it does not have an exact solution to compare with. Overall, our results are in good qualitative agreement with those found in~\cite{Del_Zanna_2003, Etienne:2010ui, Mosta:2013gwu, Neuweiler:2024jae}.

\subsection{Bondi accretion}

We now turn our attention to the first test where the GRMHD solver is coupled to a fixed curved metric background. To this end we consider the problem of spherical accretion onto a Schwarzschild black hole. This problem admits a quasi-analytic stationary solution~\cite{1984ApJ...277..296H} and is therefore ideal to test the correctness of a GRMHD solver in spacetimes that are not flat.

For our initial data, we follow Ref.~\cite{White_2016} and state the problem as follows: given a nonrotating black hole of mass $M$ surrounded by a fluid described by an adiabatic EOS with index $\Gamma$ and adiabat $K$, we define the polytropic index as $n = 1/(\Gamma-1)$ and assume that the fluid flow into the horizon is transonic. As a result, there exists a radius $r_c$ such that for $r<r_c$---where $r$ is the Schwarzschild radial coordinate---the coordinate speed of the fluid is supersonic.

From the conditions of transonicity, spherical symmetry and hydrostatic equilibrium, it is possible to find the velocity and temperature at $r_c$ to be~\cite{1984ApJ...277..296H}
\begin{align}
  & u^r_c = - \sqrt{\frac{M}{2\, r_c}} \,, \\
  & T_c = \left(\frac{n}{n+1}\right)\,
  \frac{(u^r_c)^2}{1-(n+3) (u^r_c)^2} \,.
  \label{eq:bondi_Tc}
\end{align}
The temperature of the fluid at any radius can then be found by solving the following equation
\begin{equation}
    (1+(n+1)\,T)^2 \left( 1 - \frac{2\, M}{r} + \frac{\mathcal{C}_1^2}{r^4 \,
    T^{2n}} \right) = \mathcal{C}_2 \,,
\end{equation}
where the constants $\mathcal{C}_1$ and $\mathcal{C}_2$ can be determined from the equations of baryon number and energy conservation, respectively
\begin{align}
  \mathcal{C}_1 & := T^n_c u^r_c r^2_c \,, \label{eq:bondi_C1} \\
  \mathcal{C}_2 & := (1+(n+1)\,T_c)^2 \left( 1 - \frac{2\, M}{r_c} + (u^r_c)^2
  \right) \,.
\end{align}

Once the temperature is known, the radial velocity can be found from~\eqref{eq:bondi_C1} and the rest of the fluid variables can be determined from the EOS
\begin{equation}
  \rho=\left(\frac{T}{K}\right)^n\,,
  \qquad
  p=T\,\rho\,.
\end{equation}
Following~\cite{White_2016, Kiuchi_2022, Stone2024}, we choose $M=1$, $\Gamma=4/3$, $K=1$, and $r_c=8$ so that the accretion rate in code units reads $\dot{M} = 0.085$ and the accretion time-scale $t_{\rm acc}:= M/\dot{M} \simeq 11.8$.

Given the spherically symmetric nature of the problem, a purely radial magnetic field should not affect the stationary solution. We thus initialize a magnetic field of the form
\begin{equation}
B^r = \frac{B_0}{\sqrt{\gamma} \, r^2} \,,
\end{equation}
and we note that while this expression is divergence free everywhere except at the origin in the continuous limit, its discrete divergence is not guaranteed to be zero since it cannot be initialized as the curl of a vector potential. Nonetheless, the CT scheme should maintain the stationary solution and leave the initial divergence unchanged; hereafter we refer to tests with $B_0=0.889$, such that the magnetization at the sonic and horizon radii is $(b^2/\rho)\simeq 0.46$ and $22.9$ and the corresponding plasma $\beta$ is $2 \, p/b^2\simeq 0.33$ and $1.1\times 10^{-2}$, respectively.

Since \texttt{GRACE} employs Cartesian coordinates, the solution is transformed to Kerr-Schild coordinates and evolved on the three-dimensional grid covering $(x,y,z)\in[0,16]^3$. We excise the fluid variables in all cells with $r \leq 1$, i.e.~inside the horizon located at $r=2$, setting them to a stationary and tenuous atmosphere, and we keep the outer boundary fixed at the Bondi solution. We run this simulation at two resolutions $h_{\rm LR} = 0.125$ and $h_{\rm HR} = 0.0625$ up to a time $t_{\rm final}\simeq 10\, t_{\rm acc}$.

\begin{figure}[h!tb]
\centering
\includegraphics[width=0.5\textwidth]{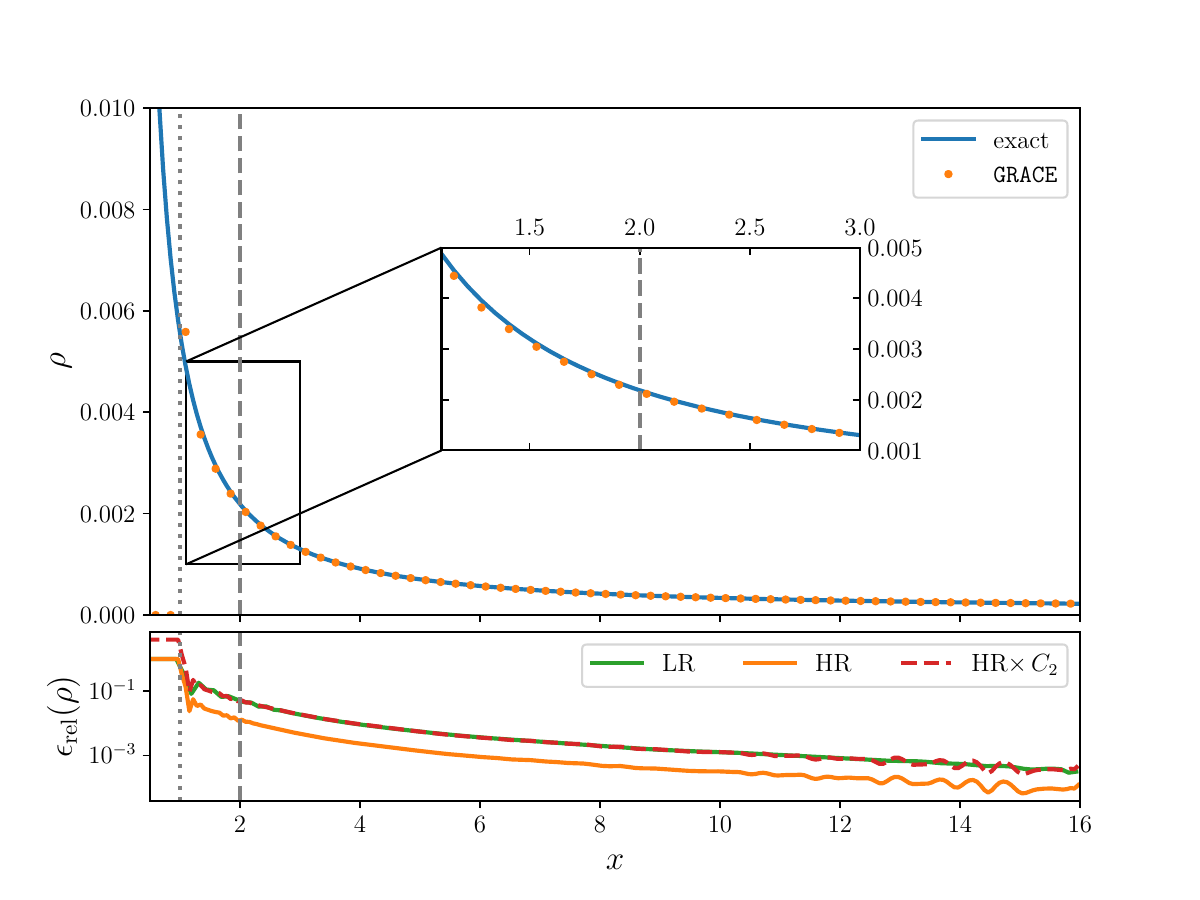}
\caption{\textit{Top:} Comparison of numerical and fiducial solutions for the rest-mass density of the fluid in the Bondi test at $t_{\rm final}\simeq 10\,t_{\rm acc}$. \textit{Bottom:} Convergence of the relative error on the density between low and high-resolution runs. In both panels the vertical dashed line indicates the location of the BH horizon, whereas the dotted line shows the boundary of the interior excision region.}
\label{fig:bondi_conv}
\end{figure}

Figure~\ref{fig:bondi_conv} shows with orange dots a one-dimensional cut of the rest-mass density profile along the $x-$axis at the end of the simulation obtained by \texttt{GRACE} and compares to the fiducial solution, shown as a blue line. The code is clearly able to preserve the solution accurately despite the excision treatment inside the horizon (the excision radius is shown as a vertical dotted line). The bottom panel of the same figure shows the relative error of the rest-mass density between the simulations at two resolutions and the fiducial solution. The red dashed line in the same plot shows the error in the high-resolution setup rescaled by the factor $C_2 := (h_{\rm LR}/h_{\rm HR})^2 = 4$ expected for second-order convergence, thus demonstrating that \texttt{GRACE} converges to the semi-analytic solution at the expected order outside the horizon.

\begin{figure}[h!tb]
\centering \includegraphics[width=0.48\textwidth]{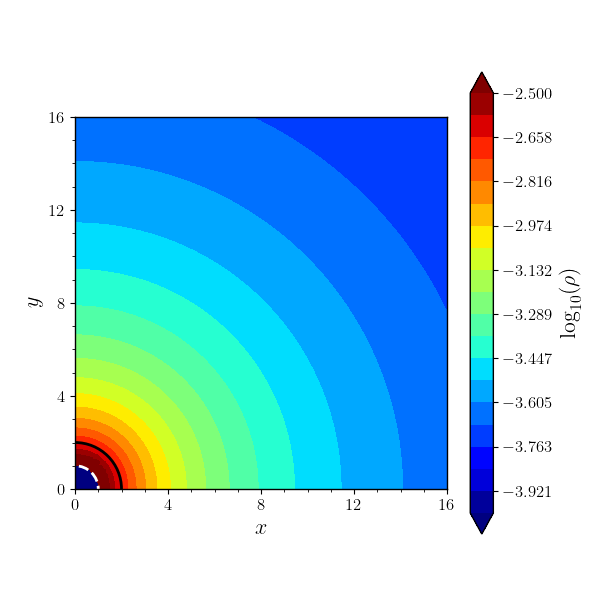}
\caption{Numerical solution for the rest-mass density in the equatorial plane for the magnetized Bondi accretion test at $t\simeq 20\, t_{\rm acc}$, with the white dashed line marking the boundary of the excised region inside the black-hole horizon, which is indicated by the black line. The circular shape of constant-density contours indicates that the evolution preserves spherical symmetry accurately, also when adopting a Cartesian coordinate system.}
\label{fig:bondi_rho}
\end{figure}

To additionally test the ability of the code to maintain the stationary solution over a longer time-scale, we extend the high-resolution simulation to $\sim 20\,t_{\rm acc}$, and report the final snapshot of the rest-mass density in the $(x,y)$ plane in Fig.~\ref{fig:bondi_rho}. As can be seen from the figure, the code maintains the spherical symmetry of the profile despite the Cartesian discretization.

\subsection{Neutron-star oscillations}

As our next test, we simulate the evolution of an equilibrium solution of the Tolman--Oppenheimer--Volkoff equations representing a nonrotating neutron star with a gravitational mass $M=1.4\, M_\odot$. The star is described by a polytropic EOS with $\Gamma=2$ and $K=100$ and has a central density of $1.28\times 10^{-3}$ in geometrized units; we will refer to it as model A0 in the following. This model has been considered numerous times in the literature and constitutes a de-facto standard test for the implementation of GRHD solvers~\cite{Font:2001ew, Radice2013c, Cook:2023bag, Fields:2024pob}.

\begin{figure*}[h!tb]
\centering
\includegraphics[width=0.48\textwidth]{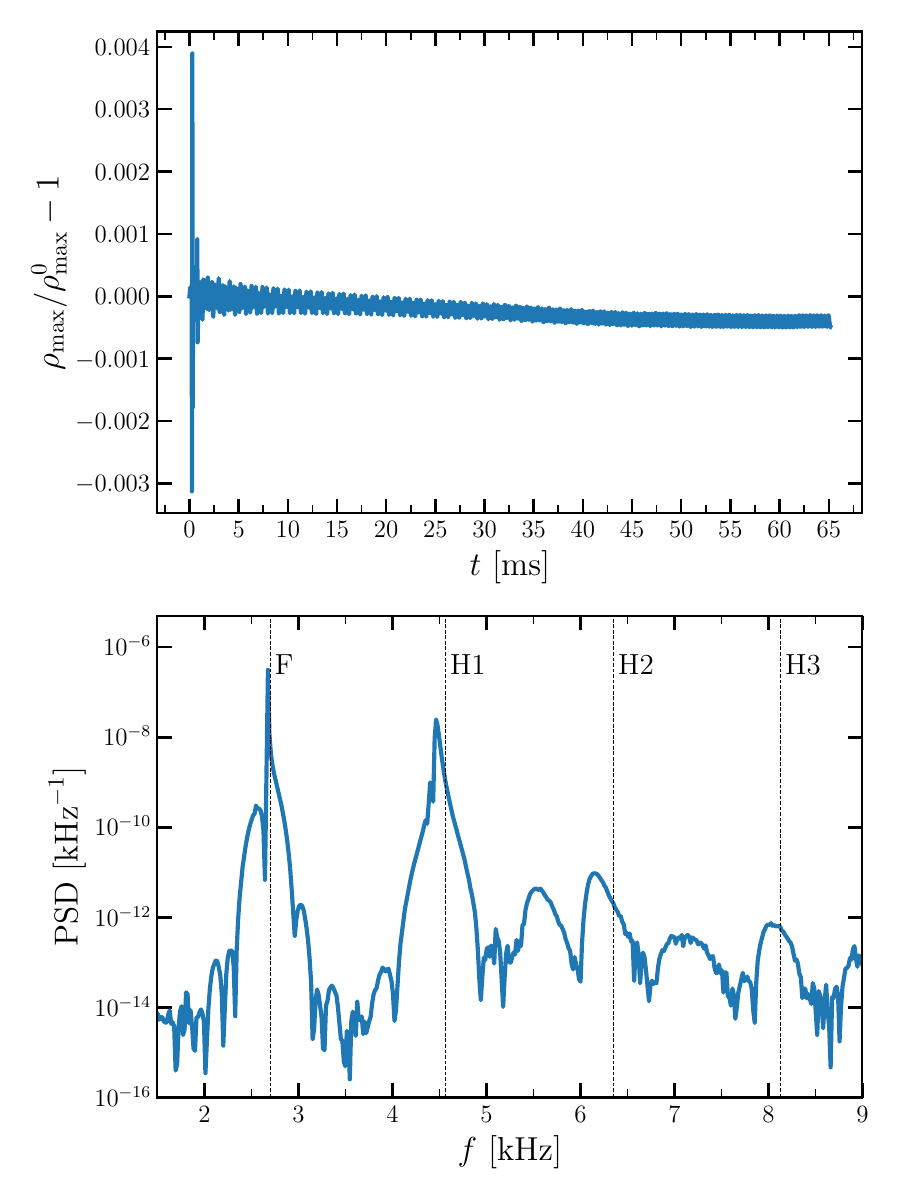}
\includegraphics[width=0.48\textwidth]{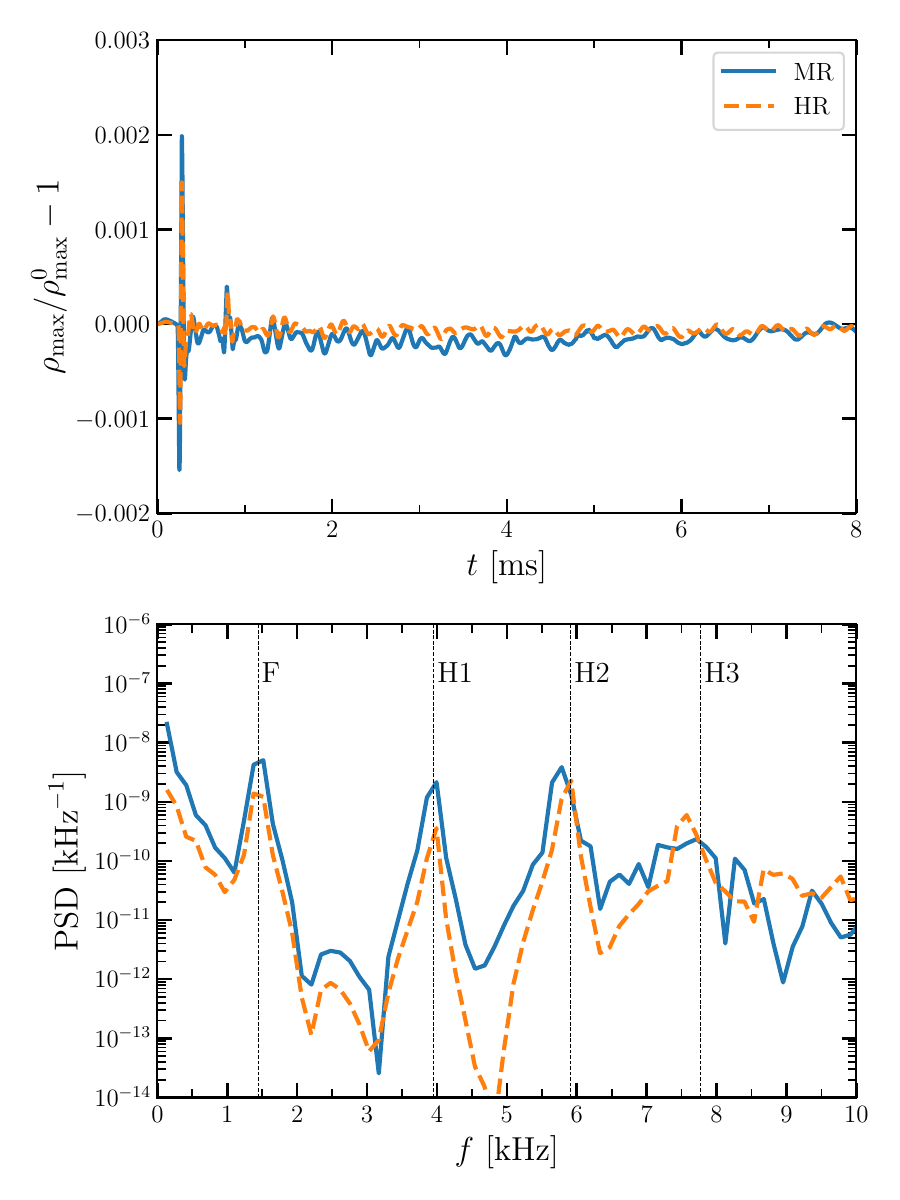}
\caption{Central-density oscillations in the TOV test (top panels) and their Fourier transform (bottom panels), where vertical lines indicate the fundamental mode and the first three overtones obtained from perturbation theory~\cite{Font:2001ew}. \textit{Left}: fixed-metric evolution in the Cowling approximation. \textit{Right}: evolution of the fully backreacted Z4c system. }
\label{fig:A0}
\end{figure*}

We first consider the evolution of this system on a fixed background metric, employing the so-called Cowling approximation, on a grid spanning $[0,128\, M_\odot]^3$---exploiting the symmetry of the solution by imposing reflection boundary conditions across the three coordinate planes, thus evolving a single octant---with three nested fixed refinement levels and finest grid spacing $h_{\rm fine} = 0.167 M_\odot$. The solution is obtained with a strong stability preserving third-order RK integrator and a fixed CFL factor of $0.2$. The evolution employs a hybrid EOS with a cold polytrope corresponding to the initial data EOS supplemented with an ideal-gas thermal part\footnote{We note that, while this EOS is equivalent to the pure ideal-gas EOS at the continuum level, in the numerical discretization of the problem this EOS enforces $\epsilon\geq \epsilon_{\rm cold}(\rho)$, which does not hold in general in the presence of numerical dissipation.} with the same $\Gamma_{\rm th}=2$. The evolution of the maximum rest-mass density, normalized by its initial value, $\rho_{\rm max}/\rho^0_{\rm max}$, together with the power spectral density (PSD) of the corresponding neutron-star oscillations triggered by numerical noise, is shown in the top and bottom left panels of Fig.~\ref{fig:A0}. Perturbative reference results from the literature~\cite{Font:2001ew} are shown for comparison, demonstrating that \texttt{GRACE} accurately reproduces the fundamental mode ($F$) and first overtone ($H1$). The other overtones $H2$, and $H3$ are less pronounced and show poorer agreement, because they are more strongly damped by finite-resolution and floor effects.

The top and bottom right panels of Fig.~\ref{fig:A0} show the evolution of the same neutron-star equilibrium configuration, now with a fully dynamical background metric evolved using the Z4c formulation. We evolve this system on the same grid $[0,128\, M_\odot]^3$, but at two different resolutions of $h_{\rm MR} = 0.125\,M_\odot$ and $h_{\rm HR}\sim 0.083\, M_\odot$. The evolution is carried out with a fourth-order RK time-stepper and a fixed CFL of $0.5$, employing the pure ideal gas EOS with $\Gamma=2$. \GRACE\ can be seen to reproduce the oscillations accurately, with both resolutions reproducing the $F$ mode frequency from perturbation theory within the discretization error of the Fourier transform. Analogously, the error on the first overtone is $-0.17\%$ for the MR setup and $0.10\%$ for the high-resolution setup. While the MR simulation does not capture the second overtone, the HR setup correctly reproduces the frequency within $-0.63\%$. Neither setup captures the third overtone faithfully. This is expected, since higher overtones of the fundamental radial mode of a TOV oscillation have more structure near the star's surface, where the numerical discretization of the problem suffers from the highest errors. Finally, we note that the medium-resolution setup shows a small but visible drift away from the initial central density, which is mitigated by increasing the resolution.

\subsection{Perturbed Spinning Puncture}

The next test setup we consider is in vacuum, i.e., does not involve the solution of the matter equations, and is therefore specifically aimed at testing the \GRACE\ solver of the Einstein equations. More specifically, and following~\cite{Daszuta:2021ecf, Zhu:2024utz}, we initialize a spinning puncture with gravitational mass $M=1$ and spin parameter $0.5$ and perturb it with a second, nonspinning puncture of mass $M_{\rm per} = 10^{-12}$ at a separation of $2\times 10^{-5}\,M$. The initial data is constructed using the open-source code \texttt{TwoPunctures}~\cite{Ansorg:2004ds}.
The grid covers $(x,y,z) \in [-1024, 1024]^2 \times [0, 1024]$ and we impose $z$-symmetry, i.e., reflection symmetry across the $(x,y)$ plane. We employ six levels of FMR so that the finest resolution is $h_{\rm fine} = 0.0833\,M$. With this setup, the second puncture is under-resolved and merges immediately with the primary, effectively acting as a perturbation. The simulation in \GRACE\ is performed with a fourth-order RK time-stepper and a CFL factor of $0.4$. Gravitational waves are extracted by integrating the projection of the Newman--Penrose scalar $\Psi_4$~\cite{Bruegmann:2006ulg} onto spin weighted-spherical harmonics on a sphere of fixed coordinate radius $R_{\rm GW}$. The grid resolution at the wave-extraction radius is $h_{\rm GW} = 0.6667\,M$.

\begin{figure}[h!tb]
\includegraphics[width=0.98\columnwidth]{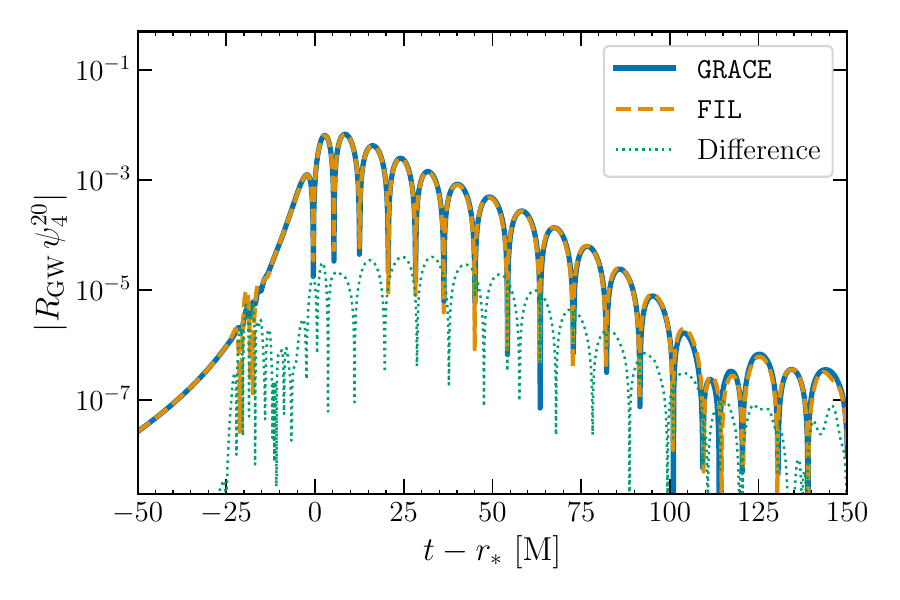}
\caption{Absolute value of the real part of the dominant ($\ell=2, m=0$) mode of the Weyl scalar $\Psi_4$ in the spinning puncture test extracted on a 2-sphere of radius $R_{\rm GW}=60\, M$. The solution obtained by \GRACE\ is compared to that of the \texttt{FIL}~code on a comparable grid structure. The difference of the results obtained by both codes is also shown as a green dotted line, highlighting the good agreement between them.}
\label{fig:spinning_puncture}
\end{figure}

As comparison, we run the same simulation with the \texttt{FIL} code~\cite{Most:2019kfe}, where---differently from \GRACE---the spacetime evolution is performed with a fourth-order finite differencing. The absolute value of the real part of the dominant $\ell=2,\, m=0$ component of $\Psi_4$ extracted from both codes at $R_{\rm GW}=60\, M$ is shown in Fig.~\ref{fig:spinning_puncture} as a function of retarded time $u = t-r_*$, where $r_* := r + 2\,M\,\log\left({r}/{2 M} - 1\right)$ is the tortoise coordinate. Despite the slight differences in the numerical methods implemented in the two codes, we find remarkable agreement in the dominant ringdown mode between the two simulations.

\subsection{Binary puncture evolution}

We now turn to simulations of compact binary systems, beginning with the orbital evolution and merger of two nonspinning punctures. This particular system has been considered numerous times in the literature~\cite{Bruegmann:2006ulg, Hilditch:2012fp, Zhu:2024utz}, and consists of two punctures of equal gravitational mass, $M=0.505$, and initial momenta $p^y = \pm 0.133$ directed along the $y$-axis.

\begin{figure}[h!tb]
\centering \includegraphics[width=1\columnwidth]{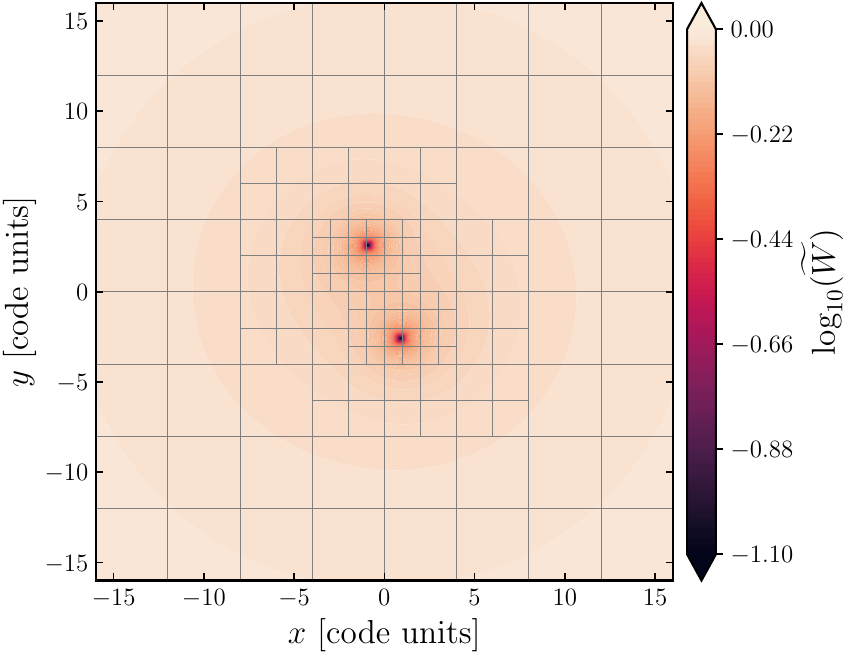}
\caption{Snapshot of the conformal factor at $t\simeq 87.5$ (in code units) showing the AMR grid structure during the late inspiral of the binary puncture system at the highest resolution considered in the main text. Each box in the figure is a quadrant, and consists of $64$ cells in each direction.}
\label{fig:bbh_Wtilde}
\end{figure}

We here employ again $z$-symmetry and set up a base grid of size $[-1024,1024]^2 \times [0,1024]$, covered by eight quadrants in the $x$- and $y$-directions and four quadrants along the $z$-axis. We then apply six levels of FMR and two levels of AMR that track the punctures. The AMR criterion is based on the minimum value of the conformal factor: a quadrant is refined whenever $\widetilde{W}_{\rm min} \leq 0.7$, and coarsened whenever the minimum over the eight quadrants sharing a common ancestor satisfies $\widetilde{W}_{\rm min} \geq 0.8$. Re-gridding happens every $50$ timesteps and is forced to respect the FMR grid, meaning that quadrants cannot be coarsened below their original level. An example of the grid structure during the inspiral of the two punctures is shown in Fig.~\ref{fig:bbh_Wtilde}, where the colormap reports the value of the conformal factor $\widetilde{W}$ in logarithmic scale.

We consider three simulations with a number of cells per quadrant that is given by $n = 32$, $48$, and $64$, corresponding to the low-, medium-, and high-resolution runs (LR, MR, and HR), corresponding to grid spacings of $h_{\rm LR}= 0.03125$, $h_{\rm MR} = 0.020833$, and $h_{\rm HR} = 0.015625$ in code units. We then evolve the system with a fourth-order RK integrator and fixed CFL factor of $0.4$. The Z4c parameters are set to $\kappa_1 = 0.02$, $\kappa_2 = 0$ and we employ a Kreiss--Oliger dissipation operator with amplitude $\epsilon_{\rm diss} = 0.5$. The Gamma-driver parameter is set to $\eta = 2$ [see Eq.~\eqref{eq:gdriver2}] and it is suppressed as $1/r$ for $r\geq 256$~\footnote{Note that we do not employ subcycling in time, so that the time-step size is the same regardless of the refinement level. For this reason we are not affected by the CFL-like instability described in~\cite{Schnetter_2010}. Despite this, we still reduce $\eta$ with radius to damp unnecessary gauge dynamics close to the boundary.}. Similarly, the coefficients $\kappa_1$ are damped as $1/r$ for $r\geq 512$ whereas the non-advective terms in the evolution of $\Theta$ are suppressed by a Gaussian factor $e^{-r^2/r_\Theta^2}$ of width $r_\Theta = 512$ (see Ref.~\cite{Kawaguchi:2015bwa}). GWs are extracted at fixed radius $R_{\rm GW}=100$.

\begin{figure}[h!tb]
\includegraphics[width=1\columnwidth]{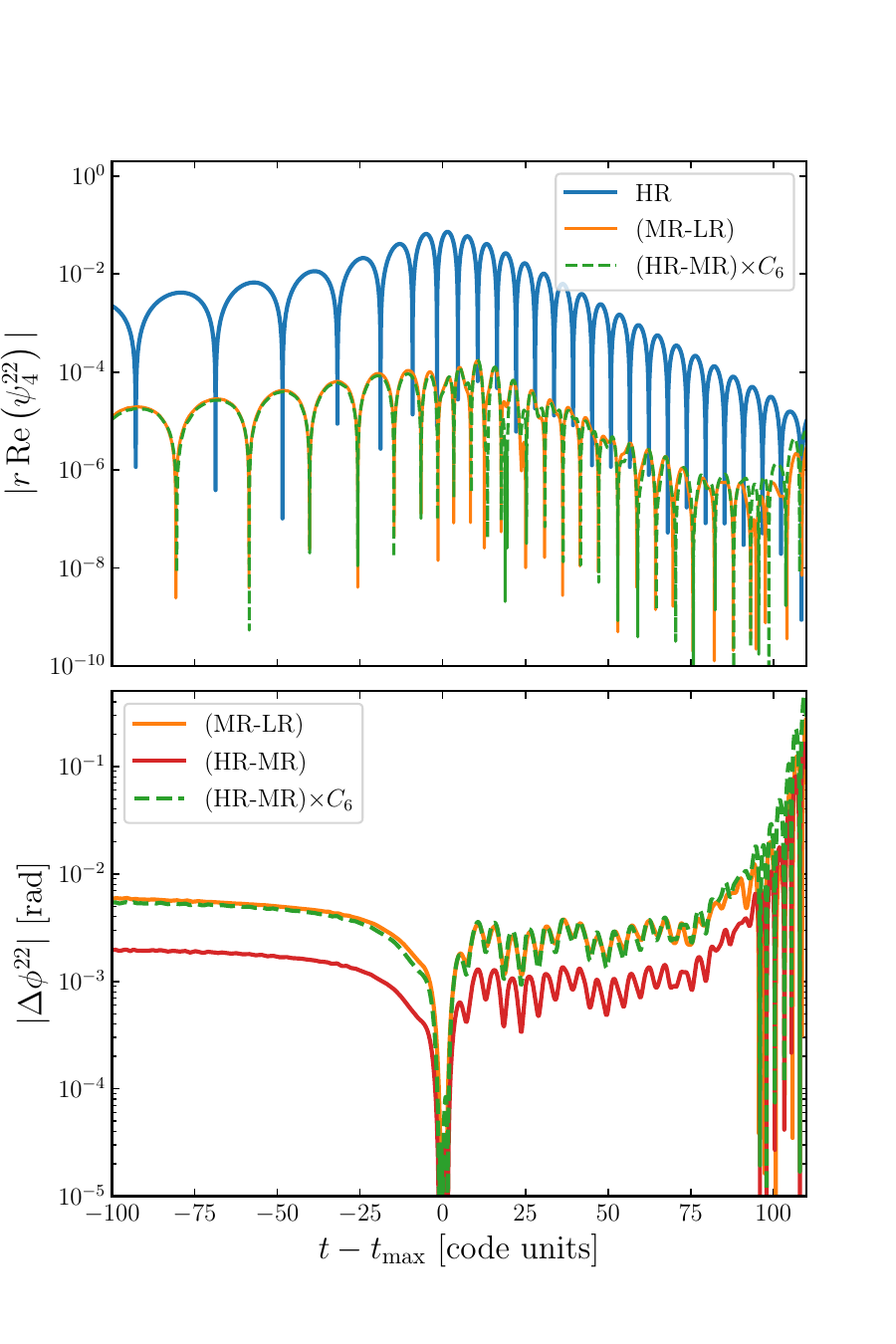}
\caption{\textit{Top:} Absolute value of the real part of the $\ell=m=2$ mode of $\Psi_4$, extracted at $r=R_{\rm GW}=100$, for the binary puncture simulation. The blue curve shows the high-resolution signal, the orange curve the medium--low resolution difference, and the green dashed curve shows the medium-to-high-resolution difference rescaled assuming sixth-order convergence. \textit{Bottom:} Absolute GW phase differences: medium--low in orange, high--medium in red, and the rescaled medium-to-high-resolution difference in dashed green.}
\label{fig:bbh_psi}
\end{figure}

The top panel of Fig.~\ref{fig:bbh_psi} shows the dominant $\ell=m=2$ mode of the $\Psi_4$ scalar, together with the differences between the solutions from \texttt{GRACE} obtained at the three resolutions considered. In particular, we define the difference in the real part of $\psi_4^{22}$ between pairs of simulations at successively higher resolutions, $h_a > h_b$, as
\begin{equation}\label{eq:delta_ab}
\Delta_{ab} := {\rm Re}\left(\psi_4^{22}(h_a)\right) - {\rm
  Re}\left(\psi_4^{22}(h_b)\right) \,.
\end{equation}
Note that the $\psi^{22}_4$ signal shown in Fig.~\ref{fig:bbh_psi} is aligned in both time and phase at $t_{\rm max}$ (i.e.,~the time where the $\psi^{22}_4$ reaches its maximum amplitude). In particular, for each simulation we extract $t_{\rm max}$ and the corresponding phase $\phi_{\rm max}$, and we subsequently shift and rotate the LR and MR signals so that they agree with the high-resolution simulation at $t_{\rm max}$. This procedure is done to ensure that the spurious, likely under-resolved phase noise coming from the initial ``junk'' radiation that reaches the detector early in the simulation does not spoil the measured convergence.

Assuming that the solution obtained by the code is the exact continuum solution of the PDE plus a truncation error that scales with the grid spacing as $h^p$, where $p$ is the formal convergence order of the scheme, we expect that the fractional residual is
\begin{equation}\label{eq:conv_order}
\frac{\Delta_{\rm LR,\,MR}}{\Delta_{\rm MR,\,HR}} \simeq \frac{(h_{\rm
    LR})^p - (h_{\rm MR})^p}{(h_{\rm MR})^p - (h_{\rm HR})^p} =: C_p \,.
\end{equation}

This particular set of simulations was performed with sixth-order accurate spatial finite differencing and fourth-order accurate time integration, so that the formal asymptotic order of the scheme is four. At finite resolution, however, the temporal truncation error carries a very small coefficient---the time step being limited by the CFL condition---while the fifth-order interpolation error at mesh-refinement boundaries is confined to lower-dimensional interfaces, so that the measured rate can be expected to be dominated by the sixth-order spatial discretization. This is confirmed by the bottom panel of Fig.~\ref{fig:bbh_psi}, which shows the pairwise GW phase differences $|\Delta\phi^{22}_{\rm MR-LR}|$ and $|\Delta\phi^{22}_{\rm HR-MR}|$, as well as the latter when scaled by the factor $C_6 \approx 12.642$, corresponding to sixth-order convergence. The rescaled error broadly agrees with the difference between the LR and MR signals, consistent with the spatial truncation error dominating the error budget at these resolutions.

\begin{figure*}[h!tb]
\centering
\includegraphics[width=0.8\textwidth]{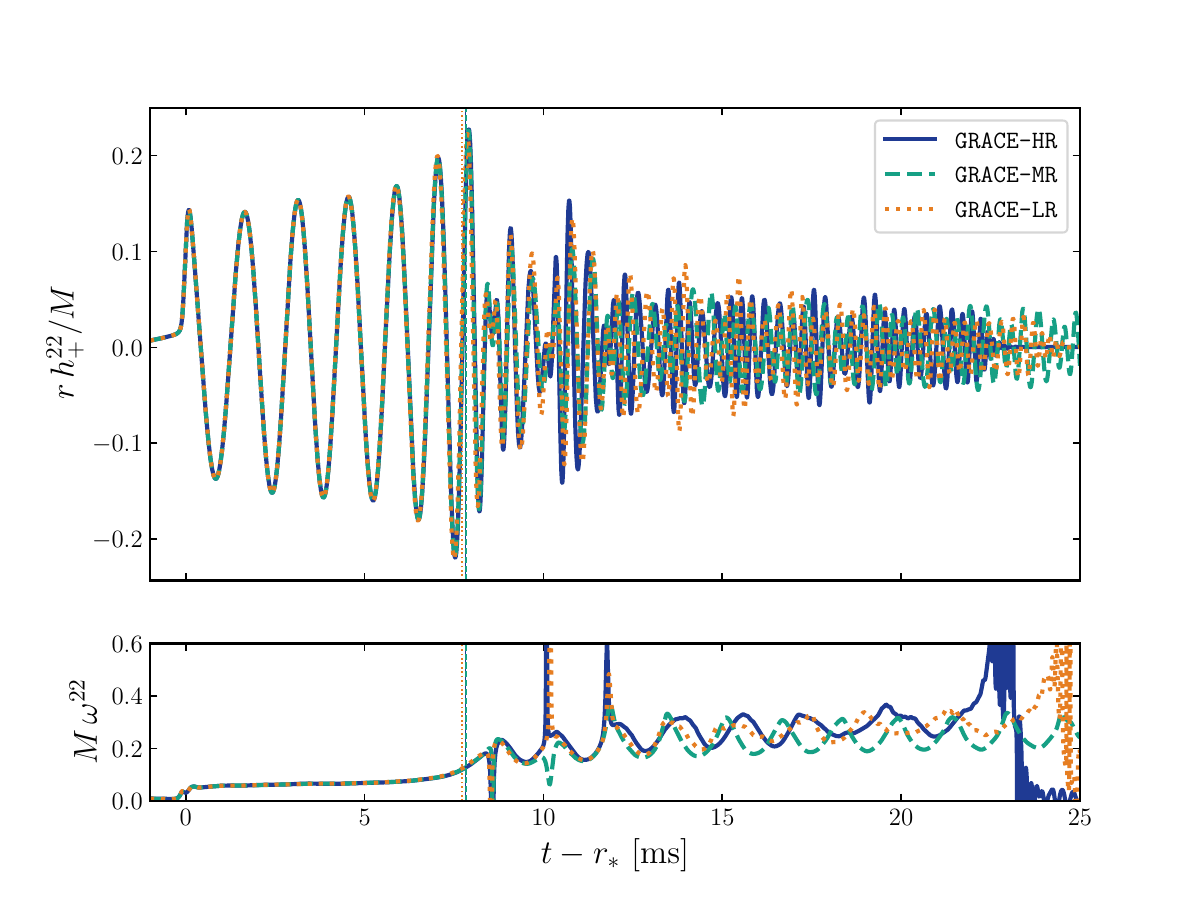}
\caption{Gravitational-wave signal from the unmagnetized $\Gamma=2$ equal-mass BNS inspiral, extracted at $R_{\rm GW}=500\,M_\odot$. \textit{Top}: real part of the $\ell=m=2$ strain mode, $r h_+^{22}/M$, as a function of retarded time, for the three resolutions LR, MR and HR. \textit{Bottom}: corresponding instantaneous angular frequency $M\omega^{22}=M\dot{\phi}^{22}$. The vertical line marks the merger time at $t_{\rm mer}\simeq 7.8\,\mathrm{ms}$. The three resolutions are visually indistinguishable throughout the inspiral, but start to disagree after merger, as expected.}
\label{fig:bns_g2_strain}
\end{figure*}

\subsection{Unmagnetized Binary Neutron-Star Inspiral}
\label{ssec:bns_g2}

As a first test of the coupled GRHD and dynamical-spacetime sectors of \GRACE\ on a compact binary, we consider the inspiral of an equal-mass, unmagnetized binary neutron-star system. The initial data are built in the irrotational quasi-equilibrium approximation with a cold polytropic EOS, $p_{\rm cold}=K\,\rho^\Gamma$ with $\Gamma=2$ and $K=123.6133$ in code units, while the time evolution is performed with an ideal-gas EOS so that shock heating during merger is captured consistently; on the initial slice the two prescriptions coincide.

This setup is by now a standard benchmark for binary neutron-star evolutions and has been extensively studied in the literature using different formulations of the Einstein equations and different numerical schemes, including by the closely related \texttt{GR-Athena++}~\cite{Cook:2023bag} and \texttt{AthenaK}~\cite{Fields:2024pob} codes, which makes it particularly well-suited to assess the consistency of \GRACE\ with the state of the art.

The initial configuration is produced with the \texttt{LORENE}~\cite{Gourgoulhon:2000nn} library and corresponds to a binary with a baryon mass of $M_b=1.625\,M_\odot$ per star, an initial coordinate separation of $45\,\mathrm{km}$, resulting in an ADM mass $M_{_{\rm ADM}}\simeq 2.99\,M_\odot$ of the binary system. The simulations are performed on a Cartesian grid with extents $[-1024,1024]\,M_\odot$ in each direction, with reflection symmetry imposed across the $(x,y)$ plane. We employ six levels of fixed mesh refinement, with the finest box covering a region of $\pm 32\,M_\odot$ around the center of mass and tracking the entire orbit up to and beyond merger. The three resolutions LR, MR and HR are obtained by varying the number of cells per quadrant on the finest level so as to achieve finest grid spacings of $h_{\rm LR}=0.25\,M_\odot \simeq 369\,\mathrm{m}$, $h_{\rm MR} = 0.1\bar{6}\,M_\odot \simeq 246\,\mathrm{m}$ and $h_{\rm HR}=0.125\,M_\odot \simeq 185\,\mathrm{m}$, respectively. The evolution uses a fourth-order RK integrator with a fixed CFL factor of $0.4$, the Z4c equations are discretized at $4{\rm th}$ order with a Kreiss--Oliger dissipation amplitude $\epsilon_{\rm diss}=0.25$, and the Gamma-driver damping parameter is set to $\eta=1.0$ (in units of $M_\odot^{-1}$) inside $r=50\,M_\odot$ and falls off as $1/r$ at larger radii. The matter evolution equations are solved with WENO-Z reconstruction and an HLLE Riemann solver, and the C2P is performed with the scheme of Ref.~\cite{Kastaun:2020uxr} without entropy backup.

The GWs are extracted on a coordinate sphere of radius $R_{\rm GW}=500\,M_\odot$ and the strain $h$ is reconstructed by twice integrating $\psi_4^{22}$ in time using the fixed-frequency-integration scheme of Ref.~\cite{Reisswig:2010di}. In particular, we use half of the initial orbital frequency as cutoff and apply a Tukey window to the $\Psi_4$ signal before the Fourier transform, suppressing spectral leakage from the finite duration of the signal.

Figure~\ref{fig:bns_g2_strain} shows the resulting real part of the dominant $\ell=m=2$ strain mode, together with the corresponding instantaneous angular frequency, $\omega^{22} := \dot{\phi}^{22}$, for the three resolutions. The signal exhibits the expected chirp morphology of a quasi-circular BNS inspiral, with approximately four orbital cycles completed before merger. The merger itself---defined as the time of maximum strain amplitude and marked by the vertical line in Fig.~\ref{fig:bns_g2_strain}---occurs at $t_{\rm merg} \simeq 7.8\,\mathrm{ms}$ for all three resolutions. After merger, the binary forms a differentially rotating hypermassive remnant~\cite{Baumgarte:1999cq} that collapses to form a black hole---defined here as the time when the minimum of the lapse first drops below $\alpha_{\rm min}=0.2$---after $\sim 14$ and $\sim 12 \,{\rm ms}$ in the low and high-resolution setups, respectively~\cite{Baiotti08}, while the medium-resolution remnant has not yet collapsed by the end of the simulation, more than $40\,{\rm ms}$ after merger; such a sensitivity of the remnant lifetime to truncation error is expected for long-lived, near-threshold remnants~\cite{Hotokezaka_2013, Koppel:2019pys}. The post-merger instantaneous angular frequency $\omega^{22}$ oscillates around $M\omega^{22}\simeq 0.2$--$0.3$, corresponding to a dominant post-merger GW frequency in the $2$--$3\,\mathrm{kHz}$ range characteristic of bar-mode oscillations of the remnant~\cite{Stergioulas_2011}. The strain signals from the three resolutions are visibly indistinguishable through the inspiral, develop a small but visible phase offset at and just after merger, and show a lack of convergence in the post-merger phase. We will comment on this in more detail below, when reviewing the convergence analysis.

\begin{figure}[h!tb]
\centering
\includegraphics[width=\columnwidth]{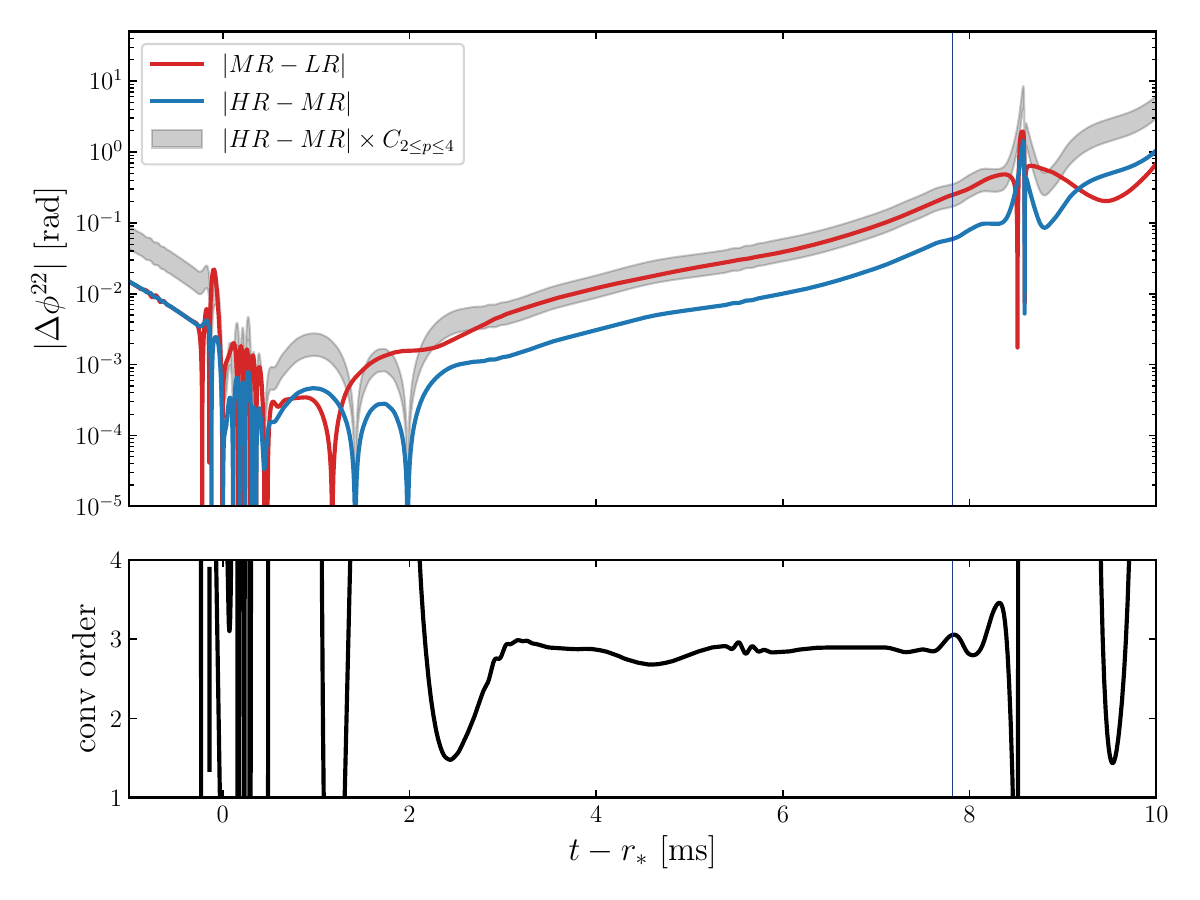}
\caption{Self-convergence of the GW phase for the unmagnetized $\Gamma=2$ equal-mass BNS inspiral as a function of retarded time; the vertical line marks the merger time. \textit{Top:} the red and blue curves are the pairwise differences $|\Delta\phi^{22}_{\rm MR-LR}|$ and $|\Delta\phi^{22}_{\rm HR-MR}|$; the gray band shows $|\Delta\phi^{22}_{\rm HR-MR}|$ rescaled by the self-convergence factor $C_p$ [Eq.~\eqref{eq:conv_order}] for $p\in[2,4]$. The coarser-pair difference remains within the band throughout the inspiral, near the $p\simeq2.8$ locus; the post-merger breakdown is expected~\cite{PhysRevD.85.104030}. \textit{Bottom:} pointwise convergence order obtained by solving Eq.~\eqref{eq:conv_order} for $p$ at each time, which remains roughly constant during the inspiral.}
\label{fig:bns_g2_phase}
\end{figure}

After computing the instantaneous phase of the dominant $\ell=m=2$ strain mode for each resolution, $\phi^{22}$, we take the absolute values of the differences between successive resolution pairs, without applying any additional time or phase alignment beyond the initial alignment at $t=0$. Figure~\ref{fig:bns_g2_phase} shows these differences, $|\Delta\phi^{22}_{\rm MR-LR}|$ (red) and $|\Delta\phi^{22}_{\rm HR-MR}|$ (blue); the gray band is the finer-pair difference rescaled by the self-convergence factor $C_p$ [Eq.~\eqref{eq:conv_order}], evaluated over $p\in[2,4]$ for the resolution triplet \{$h_{\rm LR}, h_{\rm MR}, h_{\rm HR}$\} used in this test, bracketing the range between the nominal second-order rate of the finite-volume matter sector and the fourth-order accuracy of the spacetime discretization and time integration. Rather than assigning a single convergence order, we aim at demonstrating that the code displays convergent behavior within the band dictated by the numerical methods employed: the coarser-pair difference indeed remains within this band throughout the inspiral, tracking the $p\simeq2.8$ locus; as for the SFHo binary discussed below, we regard this as indicative of where the residual sits within the band rather than as a determined convergence order. The bottom panel of Fig.~\ref{fig:bns_g2_phase} makes this explicit by reporting the pointwise order obtained by solving Eq.~\eqref{eq:conv_order} for $p$ at each instant of time: apart from the initial transient associated with the junk radiation, it remains roughly constant at this value throughout the inspiral, demonstrating that the convergence is steady rather than the result of compensating fluctuations.

That the residual sits well above the second-order edge of the band is not surprising: for the smooth flow that characterizes the inspiral, the leading second-order error terms of the matter sector carry small coefficients---with the fifth-order WENO-Z reconstruction keeping the dissipative part of the truncation error subdominant---so that, at finite resolution, the higher-order truncation errors of the spacetime sector still contribute appreciably to the measured residual. At sufficiently high resolution, the second-order flux-discretization error must eventually dominate and the effective rate settle to second order; verifying this asymptotic behavior directly would require both higher resolutions and longer inspirals, and is left to a dedicated future study. As expected, the global convergence breaks down at and after merger, as is typically observed in BNS simulations~\cite{PhysRevD.85.104030, Cook:2023bag, Fields:2024pob}.

To offer a more quantitative characterization of these simulations, which could serve as a reference benchmark for future tests of this type, we report in Tab.~\ref{tab:bns_g2_merger} the merger time $t_{\rm mer}$ (defined as the time of maximum strain amplitude and reported in physical units) and the corresponding instantaneous frequency of the dominant strain mode at merger, $f_{\rm GW}^{22}=\dot\phi^{22}/(2\pi)$, for the three \GRACE\ runs.
\begin{table}[h!tb]
\centering
\begin{tabular}{lrrr}
  \toprule
  finest     & $h_{\rm fine}$   & $t_{\rm mer}$     & $f_{\rm GW}^{22}|_{\rm mer}$ \\
  resolution & $[\mathrm{m}]$ & $[\mathrm{ms}]$ & $[\mathrm{Hz}]$ \\
  \midrule
  LR & $369$ & $7.728$ & $1335$ \\
  MR & $246$ & $7.828$ & $1381$ \\
  HR & $185$ & $7.822$ & $1367$ \\
  \bottomrule
\end{tabular}
\caption{Finest grid spacing, merger time, and merger frequency of the dominant $\ell=m=2$ mode, $f_{\rm GW}^{22}=\dot{\phi}^{22}/(2\pi)$, evaluated at peak strain amplitude for the unmagnetized $\Gamma=2$ BNS inspiral with $45\,\mathrm{km}$ initial separation at the three different resolutions shown in Fig.~\ref{fig:bns_g2_strain}.}\label{tab:bns_g2_merger}
\end{table}

The merger times all agree within $\sim 0.1\, {\rm ms}$ of each other, and the value of $\sim 7.82\, {\rm ms}$ for the highest resolution setup is in close agreement with the value $7.83\,\mathrm{ms}$ reported by Ref.~\cite{Cook:2023bag} for the same physical setup. At our highest-resolution run, whose finest grid spacing $h_{\rm HR}\simeq 185\,\mathrm{m}$ coincides with the intermediate of the three resolutions used by Ref.~\cite{Fields:2024pob} on the same physical setup, \GRACE\ obtains $f_{\rm GW}^{22}|_{\rm mer} \simeq 1367\,\mathrm{Hz}$, in agreement at the $\lesssim 1\%$ level with the value of $1359\,\mathrm{Hz}$ reported by~\cite{Fields:2024pob} at matched resolution (the merger frequencies reported there across resolutions and codes span $1329$--$1389\,\mathrm{Hz}$). We note that the merger time and $f_{\rm GW}^{22}|_{\rm merg}$ do not converge monotonically in resolution in our runs; we attribute this to numerical noise in the local time derivative of the phase near peak amplitude rather than to non-convergence of the underlying signal, as the inspiral-phase differences decrease under refinement all the way to merger (cf.~Fig.~\ref{fig:bns_g2_phase}).

\begin{figure}[h!tb]
\centering
\includegraphics[width=\columnwidth]{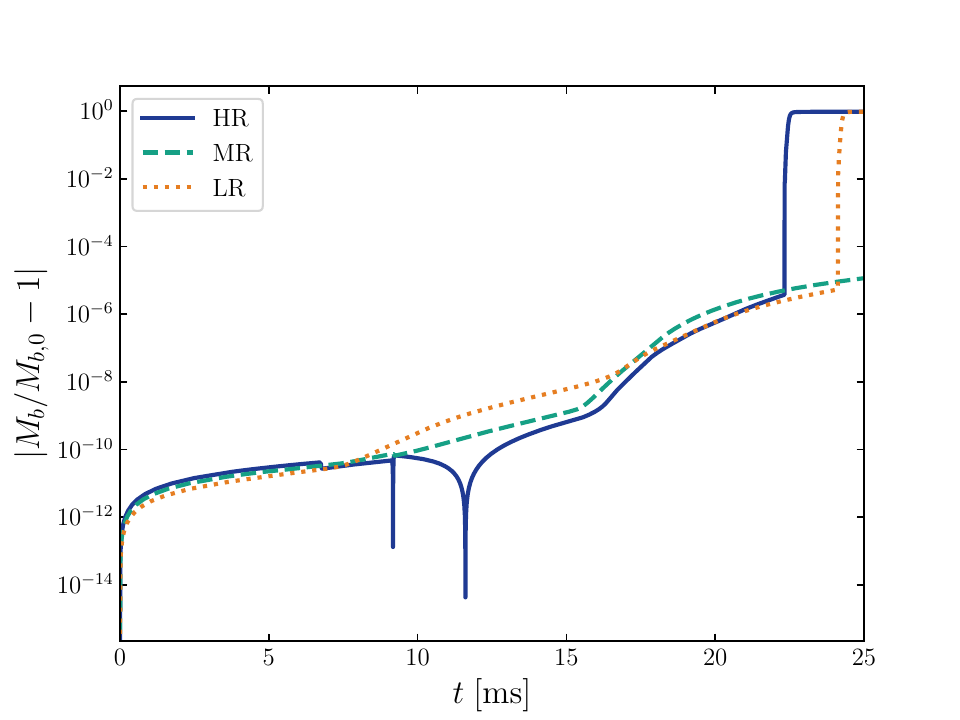}
\caption{Rest-mass conservation in the unmagnetized binary neutron star simulations performed with \GRACE. The rest mass is conserved at the round-off level, $\lesssim 10^{-10}$, through the inspiral; the subsequent slow rise is caused by tenuous matter leaving the grid, and steepens at $t\approx 16$--$17\, {\rm ms}$, when the outflow launched at merger reaches the outer boundary. The final jump to order unity marks the collapse of the remnant to a black hole.}
\label{fig:mass_cons_g2_bns}
\end{figure}

As a consistency check of the correct implementation of the conservative formulation of the matter equations and of the proper refluxing through mesh refinement interfaces in \GRACE, Fig.~\ref{fig:mass_cons_g2_bns} shows the absolute value of the relative variation of the rest-mass measured via the volume integral
\begin{equation}\label{eq:Mb}
M_b := \int \sqrt{\gamma} \, W \rho\, d^3 x \,,
\end{equation}
when compared to the total rest-mass $M_{b,0}$ on the initial spatial slice. Thanks to the robustness of the first-order flux correction~\cite{Fields:2024pob}, the code conserves rest-mass at the level of the accumulated round-off error, $\lesssim 10^{-10}$ relative to the initial value $M_{b,0}$, for the first $\sim 9$--$13\,{\rm ms}$ of evolution (increasing with resolution), after which the cumulative loss of tenuous, near-atmosphere material through the outer boundary emerges above this floor. The losses grow slowly thereafter, remaining below $\sim 10^{-8}$ until $t\approx 16$--$17\,{\rm ms}$ (increasing slightly with resolution), when the outflow launched at merger reaches the outer boundary and the mass loss steepens. Indeed, we observe that in the absence of atmosphere floorings the total mass decreases strictly monotonically, at a rate of a few parts in $10^{15}$ per timestep that is essentially independent of resolution---the signature of a systematic floating-point rounding bias in the flux-update arithmetic, rather than of truncation error, which would converge away with resolution, or of randomly accumulating round-off, which would carry no definite sign. The zero crossings in the high-resolution simulation, on the other hand, correspond to slight increases of the total mass (of the order of a few $10^{-10}\, M_\odot$) caused by cells being reset to the atmosphere floor---the mechanism recently identified in Ref.~\cite{daszuta2026exactmassconservationbinary} as the dominant source of baryon-mass violation in otherwise flux-conservative codes. In our simulations this contribution remains comparable to the round-off floor itself throughout the inspiral.

\subsection{Magnetized Binary Neutron-Star Merger}
\label{ssec:bns}

As a final and most comprehensive test of \GRACE, we consider the inspiral and merger of two magnetized neutron stars and the subsequent evolution of the remnant. To this end, we generate initial data with the open-source library \texttt{FUKA}~\cite{Grandclement:2009ju, Papenfort2021b, Tootle2021} for a binary system with chirp mass $\mathcal{M} := (M_1 M_2)^{3/5} / (M_1 + M_2)^{1/5} = 1.188\,M_\odot$ and mass ratio $q = M_1/M_2 = 0.8$, consistent with those inferred from the GW170817 event~\cite{LIGOScientific:2017vwq}. The resulting binary component masses are $M_1\approx 1.222\,M_\odot$ and $M_2\approx 1.528\,M_\odot$. The binary is initialized with the two stars on the $x-$axis at an initial distance of $45\,{\rm km}$, resulting in an initial orbital frequency of $\sim 289\,{\rm Hz}$ and a binary ADM mass of $2.72\,M_\odot$. In addition, we apply an eccentricity reduction procedure to the binary using the $3.5$ post-Newtonian estimate for the radial velocity implemented in \texttt{FUKA}~\cite{Grandclement:2009ju, Papenfort2021b, Tootle2021}.

For the microscopic matter description we employ a tabulated version of the SFHo EOS~\cite{Steiner:2012rk, Hempel_2010} from the CompOSE database. To prepare the initially cold fluid configuration of the neutron-star binary, we extract the neutrino-less beta-equilibrium EOS from the lowest-temperature slice (\ie $T=0.1\,\mathrm{MeV}$) of the full table using routines provided by our companion Python library, \texttt{GRACEpy}. We then superimpose a purely poloidal magnetic field with maximal strength $|B|_{\rm max} \sim 10^{15} \,{\rm G}$ on both stars. In particular, we define the edge-staggered vector potential $\boldsymbol{A}:=(A_{(x)~i,j-1/2,k-1/2}\,, A_{(y)~i-1/2,j,k-1/2}\,, A_{(z)~i-1/2,j-1/2,k})$, with subscripts indicating cell indices, as
\begin{equation}
\boldsymbol{A}:= \begin{pmatrix} - y_a\\ x_a\\ 0
\end{pmatrix} A^0 \max(p-p_{\rm cut},0)^n \,,
\end{equation}
where $x_a,y_a$ ($a=0,1$) are the coordinates with respect to each star's center, $p_{\rm cut}$ is chosen as $4\%$ of the maximum pressure on the initial slice and $n=2$. The amplitude $A^0$ is chosen as to obtain the desired initial field strength. The magnetic field at cell faces is then obtained as the curl of the vector potential, which, at second-order accuracy and for the $x$-component reads
\begin{align}
    (\sqrt{\gamma} \, B^x)_{i-1/2,j,k} =
  &\frac{A_{(z)~i-1/2,j+1/2,k}-A_{(z)~i-1/2,j-1/2,k}}{\Delta y}
  \nonumber\\ -&
  \frac{A_{(y)~i-1/2,j,k+1/2}-A_{(y)~i-1/2,j,k-1/2}}{\Delta z}\,,
\end{align}
(the expression for other components can be easily obtained by permutation of the indices).

To ensure that the initial magnetic field is consistent across quadrant and refinement boundaries, we compute the pressure at the cell edge as the average of cell centered values for all cells within a quadrant that touch that edge. We then apply the EMF-recirculation procedure to the resulting vector potential, ensuring that the magnetic flux computed from either side of all matching cell faces is identical at the level of round-off error.

\begin{figure}
\includegraphics[width=\columnwidth]{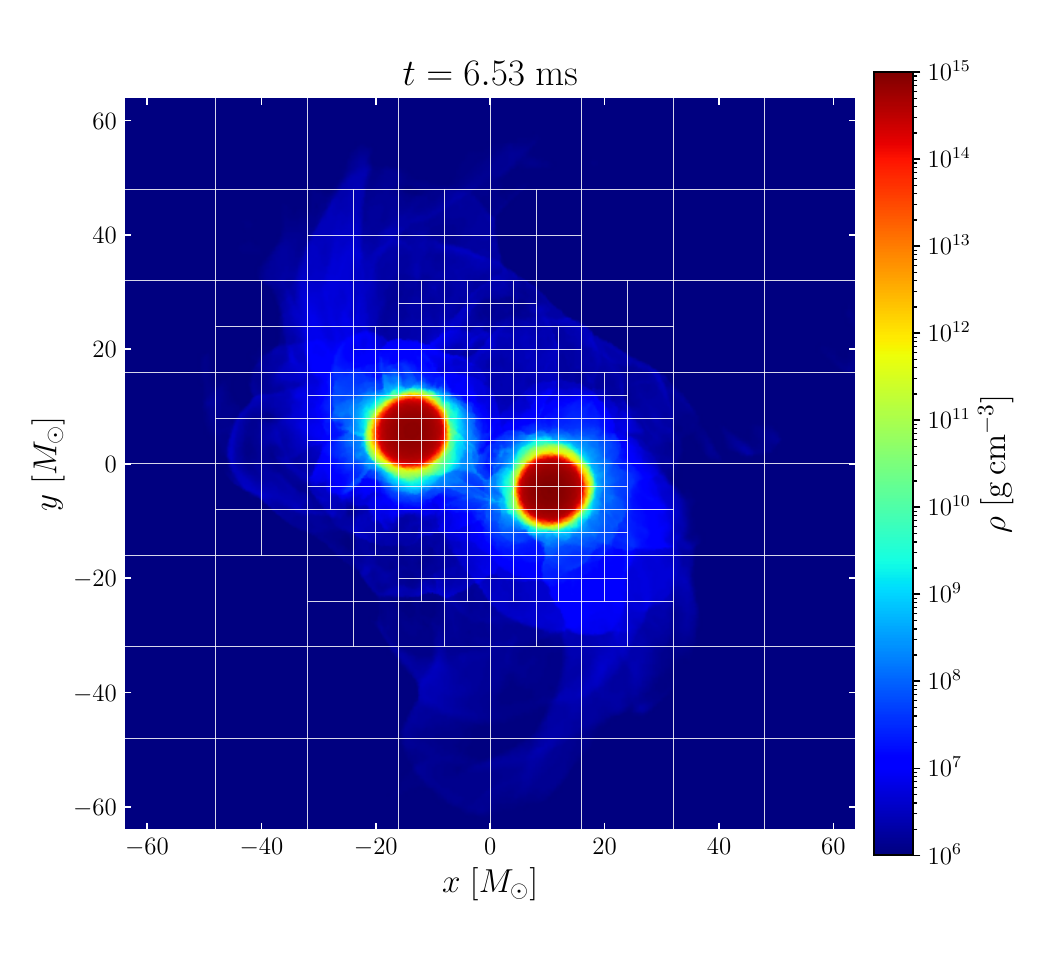}
\caption{Snapshot of rest-mass density during the inspiral in the \GRACE-HR simulation. The AMR grid structure is overlayed in white, with each square corresponding to a quadrant of $32^3$ cells.}
\label{fig:sfho_snap}
\end{figure}

The system of GRMHD equations is then evolved in \texttt{GRACE} using the fourth-order RK time-stepper with a fixed CFL factor of $0.5$ and, as in Sec.~\ref{ssec:bns_g2}, a $4{\rm th}$-order discretization of the Z4c equations. The grid has total extents $(x,y,z) \in [-1024, 1024]^2 \times [0, 1024]$ , uses four fixed refinement levels and two AMR levels following the stars. As a grid update criterion, we use the locations of the stars calculated as the center of the rest-mass integral
\begin{equation}\label{eq:xNS}
x^i_{\rm NS} := \frac{\int d^3x \sqrt{\gamma} \, W\, \rho\, x^i }{\int
  d^3x \sqrt{\gamma} \, W\, \rho} \,,
\end{equation}
where the integral extends in a fiducial region around the previously known location of the star, which we take to be a sphere of coordinate radius $r_{\rm CoM} = 8\, M_\odot$. We then flag any quadrant contained in a sphere of radius $12\, M_\odot$ around the NS centers for refinement, and any quadrant outside a sphere of radius $24 \, M_\odot$ for coarsening. When the stars plunge,  defined here as the first time their coordinate separation drops below $20\,M_\odot$, we fix the two AMR levels such that the finest level covers a region of radius $32\,M_\odot$ around the center. An example of the grid-refinement structures around the two neutron stars during the inspiral is shown in Fig.~\ref{fig:sfho_snap}, which also reports the distribution of the rest-mass density with a color map.

The base grid consists of eight quadrants per direction in the $x$ and $y$ directions, and we employ $16,24$ and $32$ cells per quadrant per direction in the low, medium and high-resolution setups, leading to finest grid spacings $h_{\rm LR}=0.25\,M_\odot \simeq 369\, {\rm m}$, $h_{\rm MR}=0.1\bar{6}\,M_\odot\sim 246\,{\rm m}$ and $h_{\rm HR}=0.125\,M_\odot \sim 185\,{\rm m}$, respectively. We set the atmosphere rest-mass density to be $\rho_{\rm atm}=10^{-14}$ in code units, and we take the corresponding temperature as $T_{\rm atm}=0.105\,{\rm MeV}$; the atmosphere tolerance $\delta_{\rm atmo}$, \ie the relative margin above $\rho_{\rm atmo}$ below which a cell is reset to atmosphere (cf.~Sec.~\ref{sec:c2p}), is set to $0.1$. Additionally, we enable FOFC and replace the wave-speed in the Rusanov Riemann solver with the speed of light in the grid frame for $\rho < 10^{-13}$. The damping parameters of the Z4c scheme are set to $\kappa_{1}=0.02$ and $\kappa_2=0$; the magnitude of Kreiss-Oliger dissipation is taken to be $\epsilon_{\rm diss}=0.25$ and the Gamma-driver damping parameter is chosen as $\eta=0.72$, damped as $1/r$ for $r>50\,M_\odot$. No neutrino emission or transport is included in these simulations.

As a useful comparison, we simulate the same binary also with the \texttt{FIL} GRMHD code~\cite{Most:2019kfe}. As discussed in previous sections, \texttt{FIL} evolves the spacetime using the Z4c formulation of the Einstein equations. The magnetohydrodynamics solver in \texttt{FIL} is an extension of the \texttt{IllinoisGRMHD} code~\cite{Etienne_2015} which supports tabulated, composition dependent EOS and employs a fourth order accurate conservative finite difference scheme based on the \texttt{ECHO} code~\cite{Del_Zanna_2007}. Differently from \texttt{GRACE}, it evolves the magnetic vector potential $A^i$ at cell edges and reconstructs the face-staggered $B^i$ from its curl, thus ensuring the divergence of the magnetic field remains zero within round-off error at cell centers. \texttt{FIL} is coupled to the \texttt{EinsteinToolkit}~\cite{EinsteinToolkit:2025_05}, and utilizes the \texttt{Carpet}~\cite{Schnetter:2003rb} code for Berger-Oliger FMR and AMR. We thus stress that there are significant differences between the MHD modules in \texttt{GRACE} and \texttt{FIL}, with the former utilizing second order finite-volume schemes and constrained transport while the latter evolves the system with fourth order accurate conservative finite differencing and a vector potential formulation of MHD.

The grid in \texttt{FIL} is taken to have the same extent as in the \texttt{GRACE} setup, and we use six levels of fixed mesh refinement to achieve a finest resolution $h_{\texttt{FIL}} = 0.1\bar{6} \, M_\odot$. We note that the simulation in \FIL\ is performed with the SFHo EOS in the \texttt{stellarcollapse}~\cite{OConnor:2009vw} tabulated format, which differs from the CompOSE table used in \GRACE\ in the parametrization of the crust EOS. Moreover, the setup employed in \FIL\ uses a different prescription for the evolution of the shift vector where advective derivatives are omitted in Eqs.~\eqref{eq:gdriver1} and~\eqref{eq:gdriver2}.

\subsubsection{Qualitative dynamics and GW emission}

We start our analysis by discussing the orbital dynamics of the binary. Using the high-resolution simulation from \GRACE, and the locations of the stars defined in Eq.~\eqref{eq:xNS}, we follow the procedure outlined in~\cite{Kyutoku_2014} and estimate the eccentricity of the initial data. From fitting $\dot{\Omega}(t)$, we measure a residual eccentricity $e\sim 8.2\times 10^{-3}$ that, while nonzero, is unlikely to significantly bias the inspiral GW signal.

\begin{figure*}
\includegraphics[width=\textwidth]{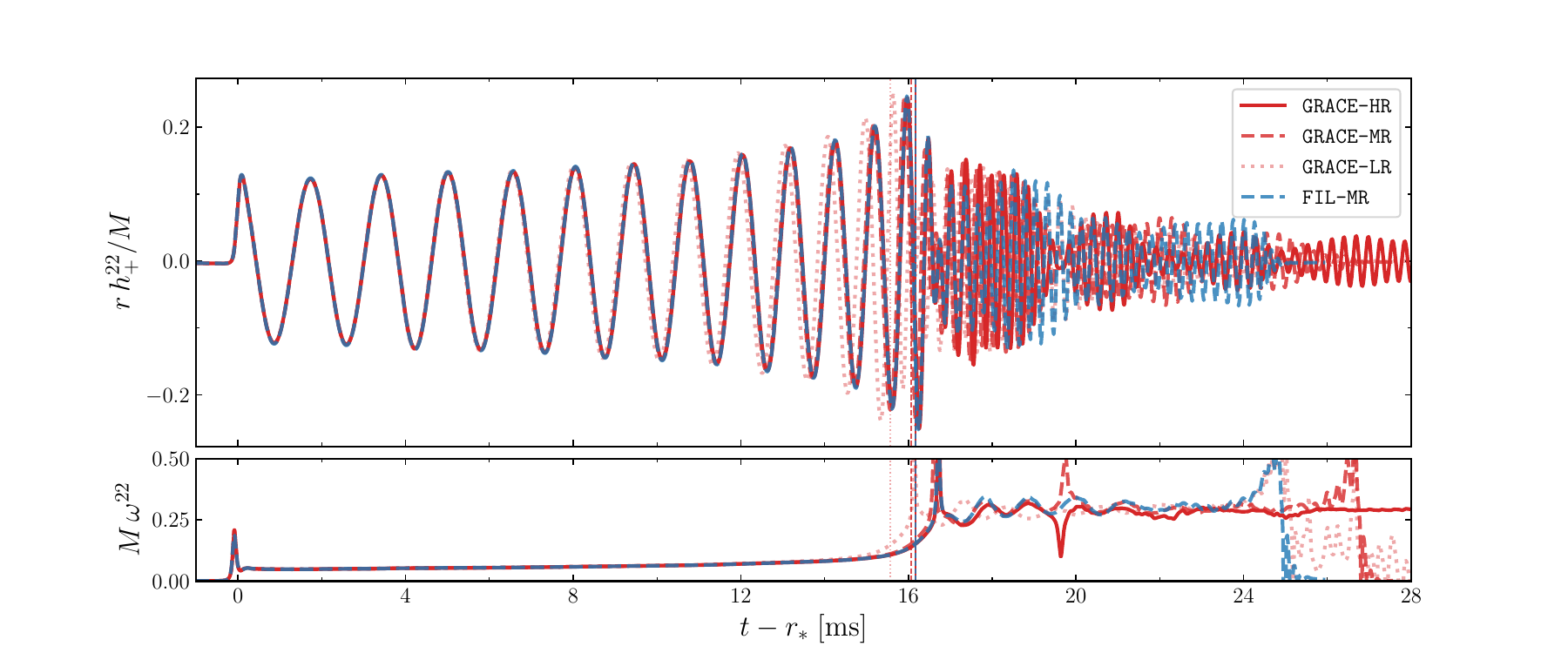}
\caption{\textit{Top:} Dominant $\ell=m=2$ strain extracted at a coordinate radius $R=500\,M_\odot$ in the BNS merger simulation with SFHo realistic EOS for the three \GRACE\ resolutions (red; HR solid, MR dashed, LR dotted) and \FIL\ (blue dashed). \textit{Bottom:} instantaneous GW frequency. The vertical lines mark the merger time of each run.}
\label{fig:sfho_gw}
\end{figure*}

Figure~\ref{fig:sfho_gw} shows the $h_+$ component of the strain as extracted at a coordinate radius $R = 500\,M_\odot$ in all of the three \GRACE\ simulations and in the \FIL\ simulation (top panel), together with the instantaneous frequency (bottom panel). The \GRACE\ simulations merge at retarded times $t-r_* \simeq 15.57\,\rm ms$, $16.07\,\rm ms$, and $16.17\,\rm ms$ for the low-, medium-, and high-resolution runs, respectively. The merger time of the \FIL\ simulation differs from the HR \GRACE\ run by less than $0.01\,\rm ms$. Using the same definition of the collapse time as in Sec.~\ref{ssec:bns_g2}, the remnant in the \GRACE\ simulations collapses to a spinning black hole $\sim 8.7$, $10.2$, and $14.7\,{\rm ms}$ after merger for the low, medium, and high resolution, respectively. The \FIL\ remnant collapses at a comparable time to \GRACE-LR. As for the binary of Sec.~\ref{ssec:bns_g2}, this large spread in collapse times reflects the sensitivity of a near-threshold hypermassive neutron star (HMNS) to small differences in truncation error.

We next proceed with describing the evolution of the magnetic fields as a result of the merger and post-merger. To this end, we follow~\cite{Neuweiler:2024jae} and define the magnetic energy in the comoving frame as
\begin{equation}
  \label{eq:e_em}
  E_{\rm EM} := \frac{1}{2} \int d^3x \, \sqrt{\gamma}\, W\, b^2 \,,
\end{equation}
and decompose it into ``toroidal'' and ``poloidal'' contributions as follows. We first define a fiducial vector along the toroidal direction $\boldsymbol{\tilde{e}}_\phi$ with components,
\begin{equation}
\tilde{e}^\mu_\phi = (0\,,-y\,,x\,,0)\,,
\end{equation}
where $x$ and $y$ are the Cartesian spatial coordinates of each point. We then project this vector onto the subspace orthogonal to the fluid 4-velocity,
\begin{equation} \label{eq:eem_proj}
\hat e^\mu_\phi = \tilde e^\nu_\phi\,(\delta^\mu_\nu + u^\mu u_\nu) \,,
\end{equation}
and normalize it as
\begin{equation}
  e^\mu_\phi := \frac{\hat e^\mu_\phi}{(g_{\nu\rho}\,\hat
  e^\nu_\phi\,\hat e^\rho_\phi)^{1/2}}\,,
\end{equation}
where the denominator is positive because $\hat e^\mu_\phi$ is spacelike by construction. After defining
\begin{equation}
  b^\mu_{\rm tor} := (b_\nu\,e^\nu_\phi)\,e^\mu_\phi\,, \qquad
  b^\mu_{\rm pol} := b^\mu - b^\mu_{\rm tor}\,,
\end{equation}
the corresponding magnetic energies follow from Eq.~\eqref{eq:e_em} with $b^2$ replaced by $b^2_{\rm tor}$ or $b^2_{\rm pol}$. The decomposition is performed in the fluid rest-frame, and the projection step in Eq.~\eqref{eq:eem_proj} is what guarantees $E_{\rm EM} = E^{\rm tor}_{\rm EM} + E^{\rm pol}_{\rm EM}$. We note that this decomposition reduces to a geometrically exact toroidal/poloidal split only in stationary, axisymmetric spacetimes.

\begin{figure}
\centering \includegraphics[width=\columnwidth]{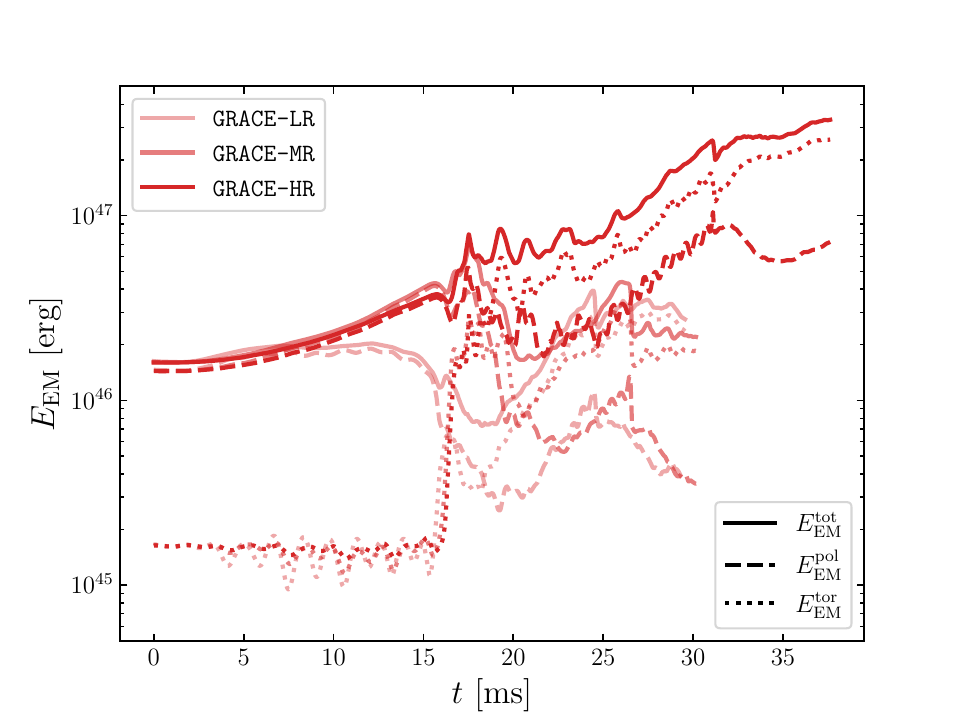}
\caption{Evolution of the comoving-frame magnetic energy $E_{\rm EM}$ [Eq.~\eqref{eq:e_em}] in the \GRACE\ SFHo BNS merger, at the LR, MR and HR resolutions. Solid, dashed and dotted lines denote the total, poloidal and toroidal contributions, respectively. The post-merger amplification is under-resolved at all three resolutions, reaching only ${\sim}10\times$ at HR (see text). }
\label{fig:grace_e_em}
\end{figure}

Figure~\ref{fig:grace_e_em} shows the evolution of the magnetic energy in the \GRACE\ BNS merger simulations. A few observations are in order. First, the LR setup is well below the resolution required to capture the small-scale dynamo at the contact interface and its post-merger magnetic-energy evolution should not be interpreted as a converged physical result. Second, even the MR and HR runs, while qualitatively reproducing the expected phenomenology, are too coarse to resolve the Kelvin–Helmholtz dynamo at merger and the magneto-rotational instability (MRI) in the post-merger disk; at HR we observe a total amplification of only ${\sim}10\times$ in the post-merger magnetic energy. Third, the post-merger growth proceeds in a series of discrete ``bounces'' rather than as a continuous climb, driven by oscillations in the poloidal component (the toroidal energy grows roughly monotonically from merger to collapse). We attribute this to numerical effects at the coarse resolutions employed here rather than to a physical feature.

Converged MHD amplification on BNS-merger scales requires resolutions of ${\sim}10$--$20\,\mathrm{m}$~\cite{Kiuchi:2015qua, Kiuchi:2023obe, Aguilera-Miret:2025nts}, about an order of magnitude finer than those employed in this study and therefore beyond the scope of this code-validation paper.

This expectation is borne out explicitly by the high-resolution magnetized-merger simulations of Kiuchi et al.~\cite{Kiuchi_2014, Kiuchi_2018}, which adopt an initial maximum field strength of $10^{15}\,$G comparable to ours and find that the magnetic-field amplification remains weak and unconverged at grid spacings of ${\sim}110$--$150\,$m, finer than the finest one employed here. A dedicated convergence study of the post-merger magnetic-field amplification with \GRACE\ at the resolutions required to capture the small-scale dynamo is left to future work.

\subsubsection{Self-convergence and cross-code comparison}
\label{ssec:comp}

We will now discuss the self-convergence properties of the \GRACE\ inspiral waveforms and compare the results with those obtained with \FIL. Since the two codes use different numerical methods, this comparison provides a meaningful and independent check of the correctness of \GRACE. Furthermore, due to the low resolution employed in this study and the lack of convergence in the post-merger phase, we will not attempt to draw comparisons between the two codes in that regime.

\begin{figure}
\includegraphics[width=\columnwidth]{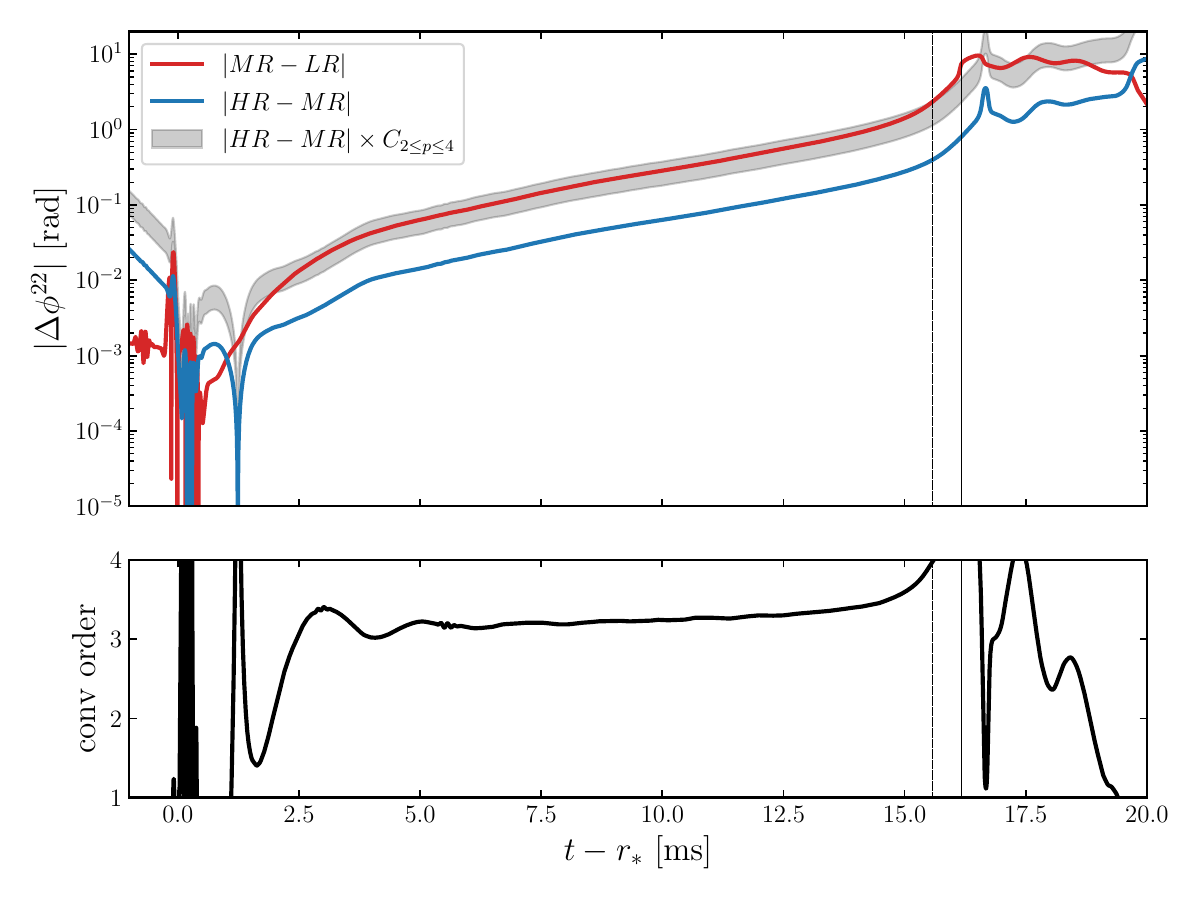}
\caption{Self-convergence of the dominant $\ell=m=2$ GW phase for the magnetized, unequal-mass SFHo binary in \GRACE, as a function of retarded time. The red and blue curves are the pairwise differences $|\Delta\phi_{\rm MR-LR}|$ and $|\Delta\phi_{\rm HR-MR}|$; the gray band shows $|\Delta\phi_{\rm HR-MR}|$ rescaled by the self-convergence factor $C_p$ [Eq.~\eqref{eq:conv_order}] over $p\in[2,4]$. The pairwise phase differences shrink under refinement and remain within the band throughout the inspiral, consistent with convergence, although the precise order cannot be reliably determined at these resolutions (see text). \textit{Bottom:} pointwise convergence order obtained by solving Eq.~\eqref{eq:conv_order} for $p$ at each time, which remains roughly constant during the inspiral. Vertical lines mark the merger times.}\label{fig:sfho_conv}
\end{figure}

Figure~\ref{fig:sfho_conv} shows the pairwise phase differences of the GWs extracted from \GRACE\ at the three resolutions, at a fixed extraction radius $R=500\,M_\odot$. As for the $\Gamma=2$ inspiral of Sec.~\ref{ssec:bns_g2}, we bracket the finer-pair difference $|\Delta\phi_{\rm HR-MR}|$ by the self-convergence factor $C_p$ over $p\in[2,4]$ (gray band) rather than assume a single order. The pairwise phase differences decrease under refinement---$|\Delta\phi_{\rm HR-MR}|$ lies below $|\Delta\phi_{\rm MR-LR}|$---and the coarser-pair difference remains within the band across the entire $\sim16\,{\rm ms}$ inspiral, with the pointwise order (bottom panel) again roughly constant, at $p\simeq3$, demonstrating convergent behavior. That it sits higher in the band than for the $\Gamma=2$ case is consistent with the pre-asymptotic character of these measurements: the effective rate reflects the superposition of truncation errors of different orders from the matter and spacetime sectors, whose balance depends on the problem and on the resolution, and the runs are too coarse---and the clean inspiral too short---to have reached the asymptotic regime, in which the rate must settle to the second order of the matter sector. As in Sec.~\ref{ssec:bns_g2}, we therefore do not quote a single convergence order. This does not weaken the validation: the errors decrease monotonically under refinement and, as we show below, the \GRACE\ waveforms agree with those of an independent code already at the level of the raw, un-extrapolated waveforms.

To obtain a robust comparison between the simulations with \GRACE\ and \FIL, we also perform a low-resolution simulation at $h_{\rm LR} = 0.25\, M_\odot$ using \FIL. We then follow the approach outlined in~\cite{Habib:2025bkb} and first align all waveforms to the high-resolution \GRACE\ result using the procedure of~\cite{Boyle_2008}, over a retarded-time window $[1000, 3200] ~ M_\odot$ (roughly $4.9\, {\rm ms}$ to $\sim 0.4 {\rm ms}$ before merger). The alignment itself is already informative, since the optimal time and phase shift for the \FIL\ simulation at medium resolution are $\Delta t\simeq 2.5\, M_\odot$ and $\Delta \phi \simeq -0.05 \, {\rm rad}$, which indicates excellent agreement of the raw waveforms. Because \FIL\ is available at only two resolutions, we cannot construct an independent \FIL\ extrapolation; instead we test whether \FIL\ approaches the \GRACE-extrapolated continuum phase as its resolution is refined. We then compute the Richardson-extrapolated phase from the \GRACE\ simulations at medium and high resolutions as
\begin{equation} \label{eq:richardson}
    \phi^{22}_{\infty} := \frac{\Delta x_{\rm MR}^p \, \phi^{22}_{\rm HR} - \Delta
      x_{\rm HR}^p \, \phi^{22}_{\rm MR}}{\Delta x_{\rm MR}^p-\Delta x_{\rm
        HR}^p} \,,
\end{equation}
where we have expressed the grid spacing as $\Delta x$ to avoid confusion with the strain. Because the self-convergence order is not sharply determined at these resolutions (Fig.~\ref{fig:sfho_conv}), we adopt $p=2$ as the reference order for the extrapolation: this is the expected asymptotic order of the scheme, and it represents a conservative choice, since a lower order attributes the largest residual truncation error to the finite-resolution waveforms. We then compare the phase from \GRACE\ and \FIL\ at each resolution against this extrapolated value.

\begin{figure*}[htb]
\includegraphics[width=0.9\textwidth]{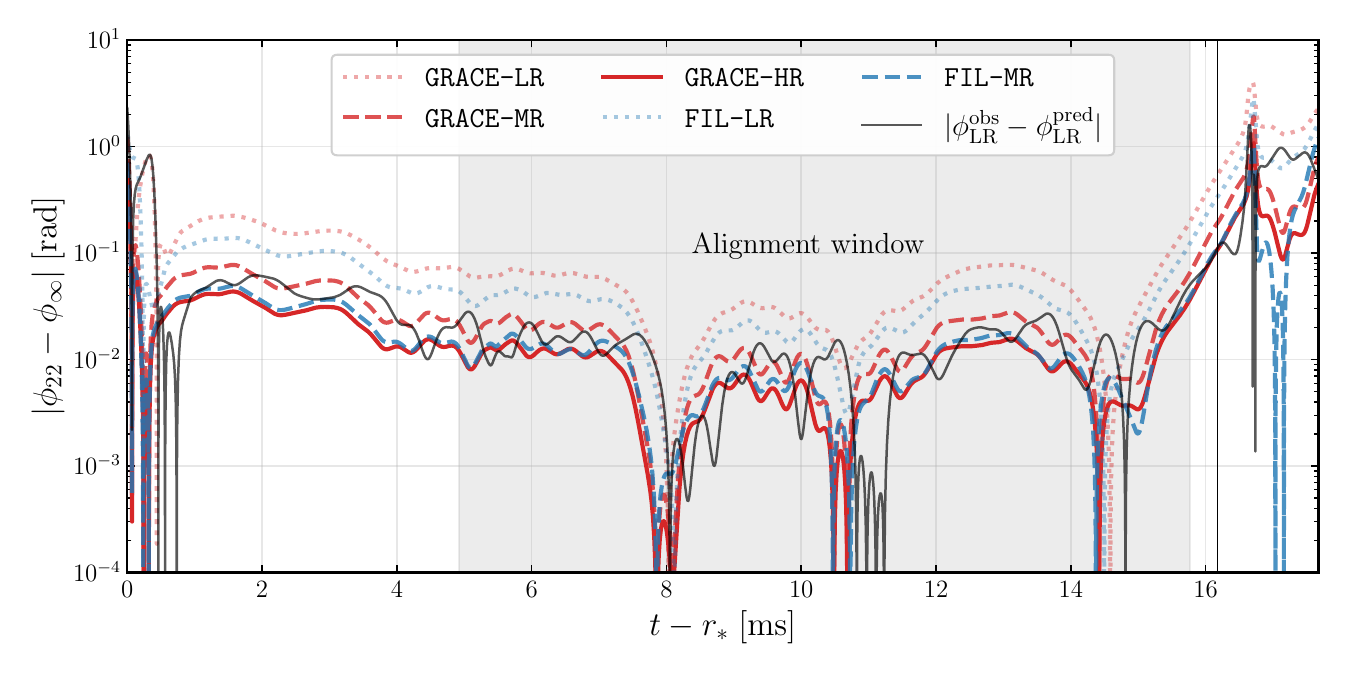}
\caption{Phase of the dominant $\ell=m=2$ mode relative to the Richardson-extrapolated \GRACE\ value $\phi_\infty$ [Eq.~\eqref{eq:richardson}, with $p=2$], as a function of retarded time, for the \GRACE\ (LR, MR, HR) and \FIL\ (LR, MR) magnetized SFHo runs; the shaded band marks the alignment window. \GRACE-HR and \FIL-MR reach comparable residuals of ${\sim}10^{-2}$--$10^{-1}\,$rad, consistent with both codes converging toward a common waveform. The black curve shows the difference between the \GRACE-LR phase and the value predicted by the Richardson model. }\label{fig:grace_fil_comp}
\end{figure*}

Figure~\ref{fig:grace_fil_comp} shows that the residuals $|\phi^{22}-\phi^{}_{\infty}|$ from both codes decrease monotonically as their respective resolutions are refined, with \GRACE-HR and \FIL-MR reaching comparable residuals of $\sim 10^{-2}-10^{-1}\,{\rm rad}$ in the inspiral window. This is consistent with both codes converging toward a common physical waveform within their respective discretization uncertainties, and suggests that any systematic difference between the two codes is at most of the order of these residuals.

As an additional self-consistency check, we plot in Fig.~\ref{fig:grace_fil_comp} also the difference between the \GRACE-LR phase and the LR phase predicted by the Richardson model. The residual remains below $\sim 10^{-1}\,{\rm rad}$ over the full window, confirming that the Richardson model with the adopted reference order is self-consistent across all three \GRACE\ resolutions. Also worth noting is that the residuals reported here are comparable in magnitude to those obtained in Ref.~\cite{Habib:2025bkb}, even though our analysis adopts a simpler finite-radius waveform-extraction procedure rather than the Cauchy-characteristic extraction to future null infinity employed there. This suggests that the residual \GRACE–\FIL\ phase difference we observe is dominated by the intrinsic discretization errors of the two codes at these resolutions and not by finite-radius extraction artifacts.

\subsubsection{Mass and divergence-free constraint conservation}

\begin{figure}
\centering
\includegraphics[width=\columnwidth]{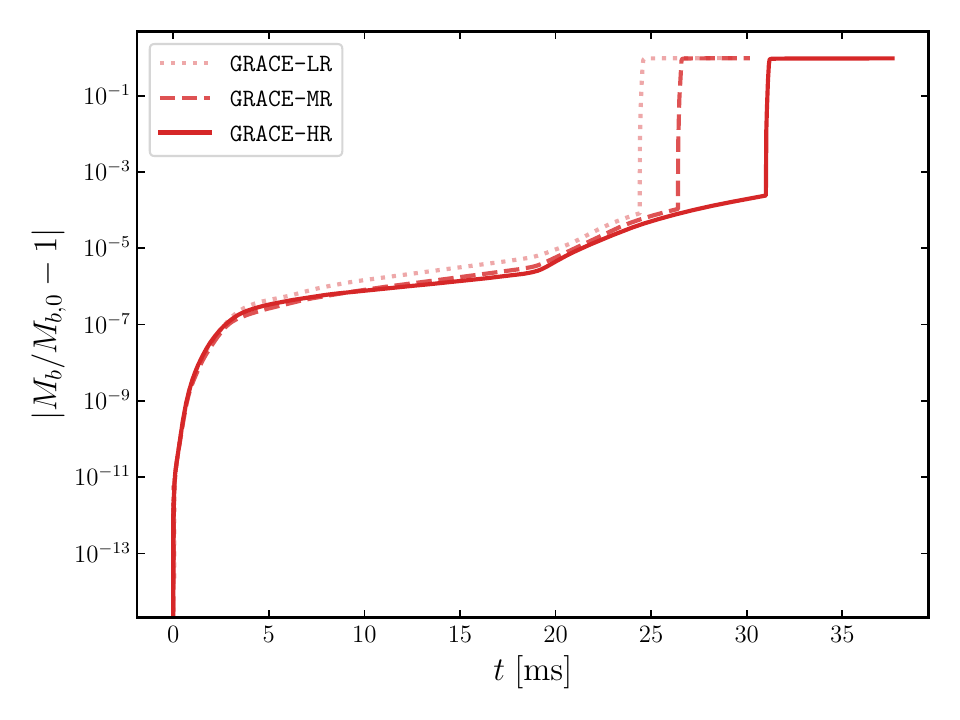}
\caption{Rest-mass conservation in the magnetized SFHo BNS simulations performed with \GRACE\ at the three resolutions considered. The rest mass is conserved to within a few parts in $10^{6}$ up to $t\approx 18\,{\rm ms}$, when shock-heated post-merger material starts to interact with the atmosphere prescription; the outflow launched at merger reaches the outer boundary at $t\approx 23$--$26\,{\rm ms}$, depending on resolution, and the final jump to order unity marks the collapse of the HMNS to a black hole.}
\label{fig:mb_cons_sfho}
\end{figure}

In Fig.~\ref{fig:mb_cons_sfho} we report the absolute value of the relative variation of the rest-mass given by Eq.~\eqref{eq:Mb}, with respect to its initial value for the three resolutions considered. Throughout the inspiral, merger, and early post-merger, that is, up to $t \approx 18\,\mathrm{ms}$, the rest-mass is conserved with a relative error of a few parts in $10^{6}$. At later times the error grows as shock-heated, low-density material increasingly interacts with the atmosphere prescription, and the outflow launched at merger crosses the outer boundary of the domain at $t\approx 23$--$26\,\mathrm{ms}$, depending on resolution. Similar to the simpler BNS merger setup of Fig.~\ref{fig:mass_cons_g2_bns}, the code maintains excellent mass conservation at all three resolutions considered, despite the higher atmosphere density. The larger relative error compared to the simpler $\Gamma=2$ ideal-gas setup is expected due to the more involved tabulated EOS and the corresponding atmosphere treatment. As in Fig.~\ref{fig:mass_cons_g2_bns}, the sudden jump to a value close to unity at later times is caused by the collapse of the HMNS to a black hole.

\begin{figure}
\centering
\includegraphics[width=\columnwidth]{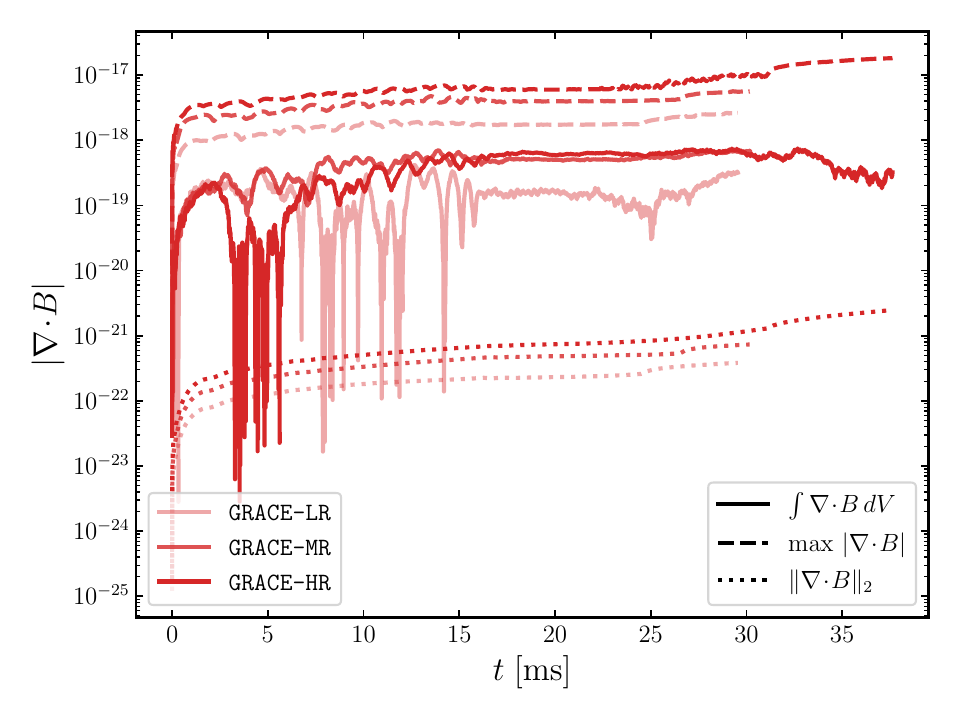}
\caption{Magnetic field divergence diagnostics in the magnetized BNS system simulated with \GRACE. For each simulation we show the $L_\infty$ norm (dashed lines), the $L_2$ norm (dotted lines) and the volume integral (solid lines). All three metrics remain within the range expected for round-off throughout the simulation.}
\label{fig:bdiv_bns_sfho}
\end{figure}

Finally, Fig.~\ref{fig:bdiv_bns_sfho} shows three measures of the magnetic-field divergence throughout the simulation, namely, the volume integral of the divergence $\int \nabla \cdot \boldsymbol{B}\, \mathrm{d}V$ (solid lines), the $L_\infty$ norm of the absolute divergence $|\nabla \cdot \boldsymbol{B}|$ (dashed lines), and its $L_2$ norm (dotted lines). All quantities are measured over the entire computational domain and different levels of transparency refer to different resolutions.

When considering the various measurements individually, it is clear that all the cell-wise diagnostics remain at the round-off floor of the discrete divergence operator for the full evolution, with the HR/LR ratio of the $L_2$ norms equal to $2.80$. This value matches the expected scaling for purely round-off-limited behavior: at each step, the discrete divergence carries a contamination of the order $\sim \varepsilon |\boldsymbol{B}|/\Delta x$ (where $\varepsilon$ is the machine round-off floor)  doubling at finer resolution, while random-walk accumulation over the factor-two larger number of timesteps contributes an additional $\sqrt{2}$, predicting a ratio of $\approx\!2.8$, which is in excellent agreement with what is observed.

Similarly, the signed volume integral of the magnetic-field divergence remains at the $10^{-19}$ level at all three resolutions with no secular drift across the full duration of the simulations. Analytically this quantity is identically zero, but it is not so when imposed numerically. This is because
%
%
when writing the cell-wise finite-difference divergences as signed sums of face fluxes, the discrete volume integral telescopes into a single sum of magnetic fluxes through every face in the grid: interior faces, shared by two neighboring cells, enter their two divergence stencils with opposite signs and cancel pairwise, leaving in principle only the outer-boundary contribution. This cancellation is exact only if the discrete $\boldsymbol{B}$ on every shared interior face has a single, common value when seen from either side.  On a uniform patch, this condition holds automatically because the staggered magnetic field has a single memory representation per face, but nontrivial conditions arise at quadrant interfaces. As already mentioned in Sec.~\ref{sec:CT}, across MPI subdomain boundaries, the shared face values must be synchronized exactly, whereas across refinement boundaries the coarse face value must equal the area-weighted sum of the four fine sub-face values it covers, and this equality must be preserved by the time update. In turn, this requires that the line integral of the electric field around the coarse face used to evolve $\boldsymbol{B}_{\rm c}$ be identical to the sum of the four line integrals around the fine sub-faces used to evolve the $\boldsymbol{B}_{{\rm f},k}$, which reduces to the condition that the coarse EMF along each shared refinement edge equal the appropriate average of the two fine EMFs along the same edge.  This is enforced in \GRACE\ by the EMF re-circulation procedure and any mismatch in this condition would result in a ``ghost'' magnetic monopole on the refinement interface that contribute to the signed volume integral while leaving each adjacent cell's individual $\nabla \cdot \boldsymbol{B}$ unchanged. Because of this, the signed integral represents a stringent test of several aspects: of the initial vector-potential discretization (which sets $\nabla \cdot \boldsymbol{B} = 0$ at $t = 0$ by computing $\boldsymbol{B}$ as the rotor of $\boldsymbol{A}$), of the MPI ghost-cell synchronization, and of the EMF refluxing at AMR refinement boundaries; none of these aspects are tested by any cell-wise diagnostic. Hence, the results reported in Fig.~\ref{fig:bdiv_bns_sfho} testify to the correct and robust implementation of the CT condition in \GRACE.

\section{Performance}
\label{sec:perf}

In this section we characterize the performance of \GRACE\ in three different regimes: single-device throughput across multiple GPU architectures, strong scaling at fixed problem size, and weak scaling at fixed work per device. We perform tests in the Cowling approximation as well as with a full dynamical spacetime, and in all cases we take as a baseline the A0 nonrotating stellar model considered in Sec.~\ref{sec:tests}. For the unigrid tests, we take a grid consisting of $64$ quadrants, each containing $64^3$ cells plus ghost-zones. For the FMR tests we set up the same TOV with $6$ levels of mesh refinement, again with quadrants of $64^3$ cells. FOFC is switched off, and the evolution is carried out with the SSPRK3 time-stepper and with WENO-Z reconstruction. Finite differencing in the Z4c equations is performed with fourth-order accurate stencils and we employ four ghost-zones. We run all simulations for $50$ timesteps, disable all outputs except for scalar reductions and measure the performance directly in the code in zone-cycles (i.e.,~cell updates) per second (zcps).

\subsection{Single-device throughput}
\label{ssec:single_device}

We begin our discussion of the performance by measuring the peak single-device throughput of \GRACE\ on three representative architectures: the AMD MI300A APU from the Viper cluster, the NVIDIA A100 GPU from the Raven cluster, and the Intel Xeon Gold 6148 CPU from the Sakura cluster (which consists of $20$ physical cores); all three clusters are hosted at the Max-Planck Computing and Data Facility. The results are shown in Tab.~\ref{tab:zcps}. We note that while the MI300A and Intel CPU share the exact same grid setup, we reduced the number of refinement levels to five on the A100 to allow the data to fit on a single GPU. Overall, the results are broadly consistent with performance of other GPU-enabled numerical-relativity codes on similar setups~\cite{Fields:2024pob}. Interestingly, we found that certain kernel-level optimizations that would improve performance on discrete GPU systems such as the A100 lead to a decrease in performance on the MI300A. Since the latter is a newer architecture, we have chosen to maintain a single, portable version of the code: the one that performs best there. Nonetheless, the ratio of throughput on different devices tracks the expected speedup coming from memory bandwidth, consistent with the code making efficient use of the more performant hardware.

\begin{table}[h!tb]
\centering
\begin{tabular}{lcc}
  \toprule Device & Cowling evol. & Full evol. \\
                  & [$10^6$ zcps] & [$10^6$ zcps] \\
  \midrule 
  {MI300A} & $24.8$ & $16.8$ \\
  {A100} & $10.5$ & $6.7$ \\
  {Intel Xeon Gold 6148} & $0.33$ & $0.18$ \\
  \bottomrule
\end{tabular}
\caption{Single-device throughput of \GRACE\ in $10^6$ zcps on the FMR TOV setup. The setup consists of the A0 nonrotating relativistic star of Sec.~\ref{sec:tests} on a fixed or evolved background metric; on a grid consisting of $6$ refinement levels for all devices except for the A100, where only $5$ levels are employed.}
\label{tab:zcps}
\end{table}

\subsection{Strong scaling}
\label{ssec:strong_scaling}

\begin{figure}[h!tb]
\centering
\includegraphics[width=.8\columnwidth]{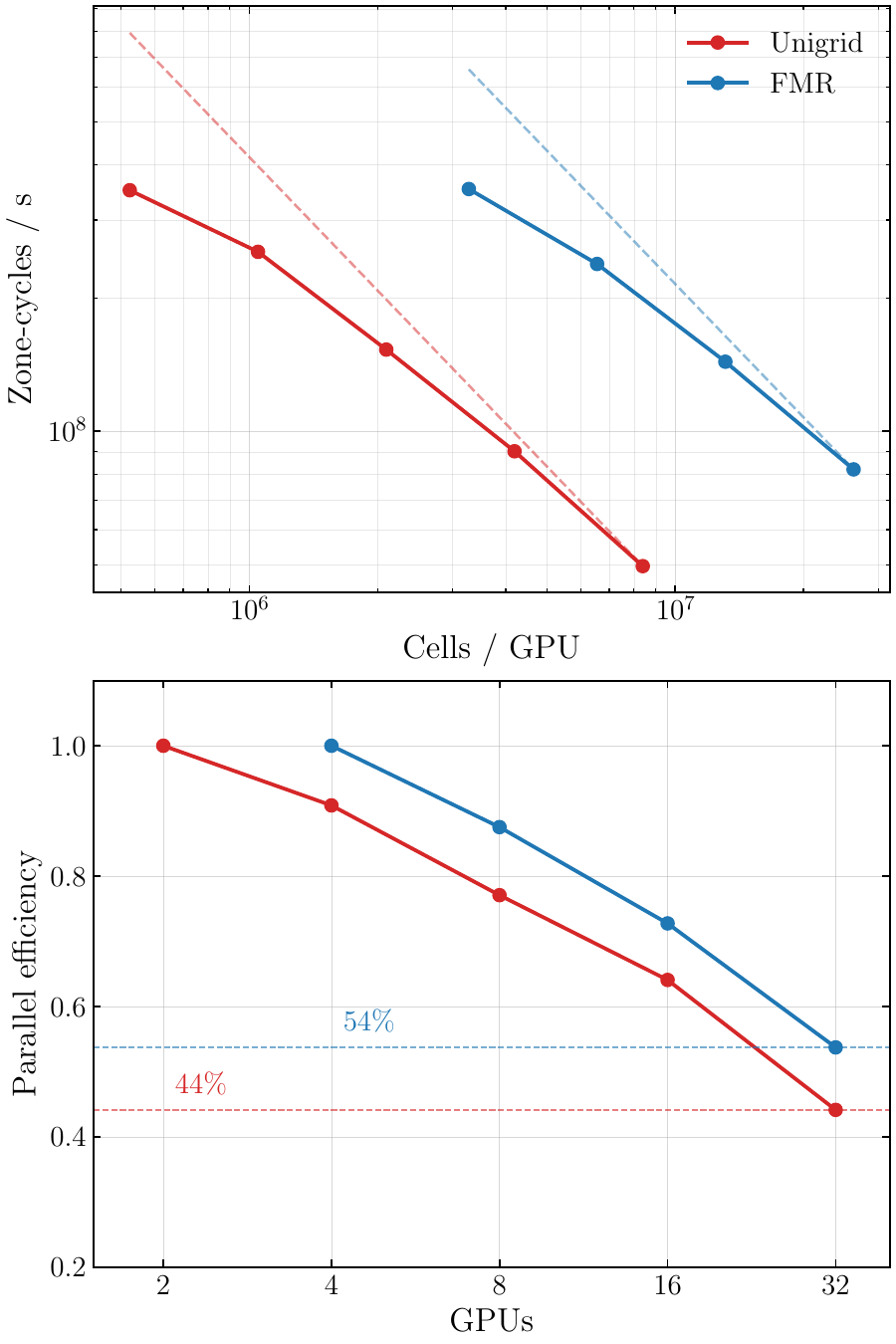}
\caption{Strong scaling of \GRACE\ on the Viper cluster for the uniform-grid (red) and six-level-FMR (blue) A0 TOV configurations. \textit{Top:} zone-cycles per second as a function of the number of cells per GPU; dashed lines indicate ideal scaling. \textit{Bottom:} parallel efficiency versus device count, normalized to the single-device throughput. The efficiency retained at the largest device counts is ${\sim}\,44\%$ for the uniform grid and ${\sim}\,54\%$ for the FMR setup (horizontal lines).}\label{fig:strong_scaling}
\end{figure}

We next probe the strong-scaling behavior of \GRACE\ by holding the problem size fixed and increasing the number of MI300A Accelerated Processing Units (APUs). Figure~\ref{fig:strong_scaling} shows the number of zone-cycles per second as a function of the number of cells per GPU (top panel) and resulting parallel efficiency as a function of device count when normalized to the single-device throughput (bottom panel). \GRACE\ maintains $\gtrsim 44\%$ efficiency on the uniform grid (red filled circles) and $\sim 54\%$ on the FMR grid (blue filled circles), with the total number of resources increasing by a factor of $16$ and $8$ for the two setups, respectively.

Two points are worth emphasizing when interpreting these numbers. First, GPUs are intrinsically poor strong-scaling targets: their performance relies on launching enough wavefronts per kernel to hide memory-access latency, and the per-device cell count drops with the device count, eventually leaving each kernel under-occupied. Second, kernel-level profiling shows that in the unigrid case the dominant bottleneck at high device count is not MPI but the physical-boundary kernels. This is because boundary loops on faces are inherently two-dimensional, on edges one-dimensional, and on corners zero-dimensional, so their surface-to-volume cost increases steeply as quadrants are distributed across more devices. Finally, the FMR case adds the further difficulty that high-order Lagrange prolongation of the metric variables has large stencils, whose dependencies serialize the ghost-exchange task graph until enough work is available downstream to hide them.

\subsection{Weak scaling}
\label{ssec:weak_scaling}

The complementary regime---which is more representative of how GPU codes are deployed at scale, and the operating point for production runs such as the BNS merger of Sec.~\ref{ssec:bns}---is weak scaling, in which the number of cells per device is held fixed as the device count increases. We construct the weak-scaling sweep by stacking identical blocks of quadrants along one coordinate direction at a time as the resource count doubles. Naturally, this grid setup is not representative of real scientific simulations. However maintaining the cell count per GPU exactly identical as we scale allows us to clearly separate potential inefficiencies coming from the code infrastructure itself from load-imbalance issues. These benchmarks also deliberately exclude file I/O, checkpointing, adaptive regridding and the interfacing with realistic tabulated EOSs in order to isolate the infrastructure efficiency; a complementary, fully realistic indication of the code's cost with all of these active---albeit not a controlled scaling measurement---is provided by the production BNS runs of Sec.~\ref{ssec:bns}.

\begin{figure}[h!tb]
\centering \includegraphics[width=.8\columnwidth]{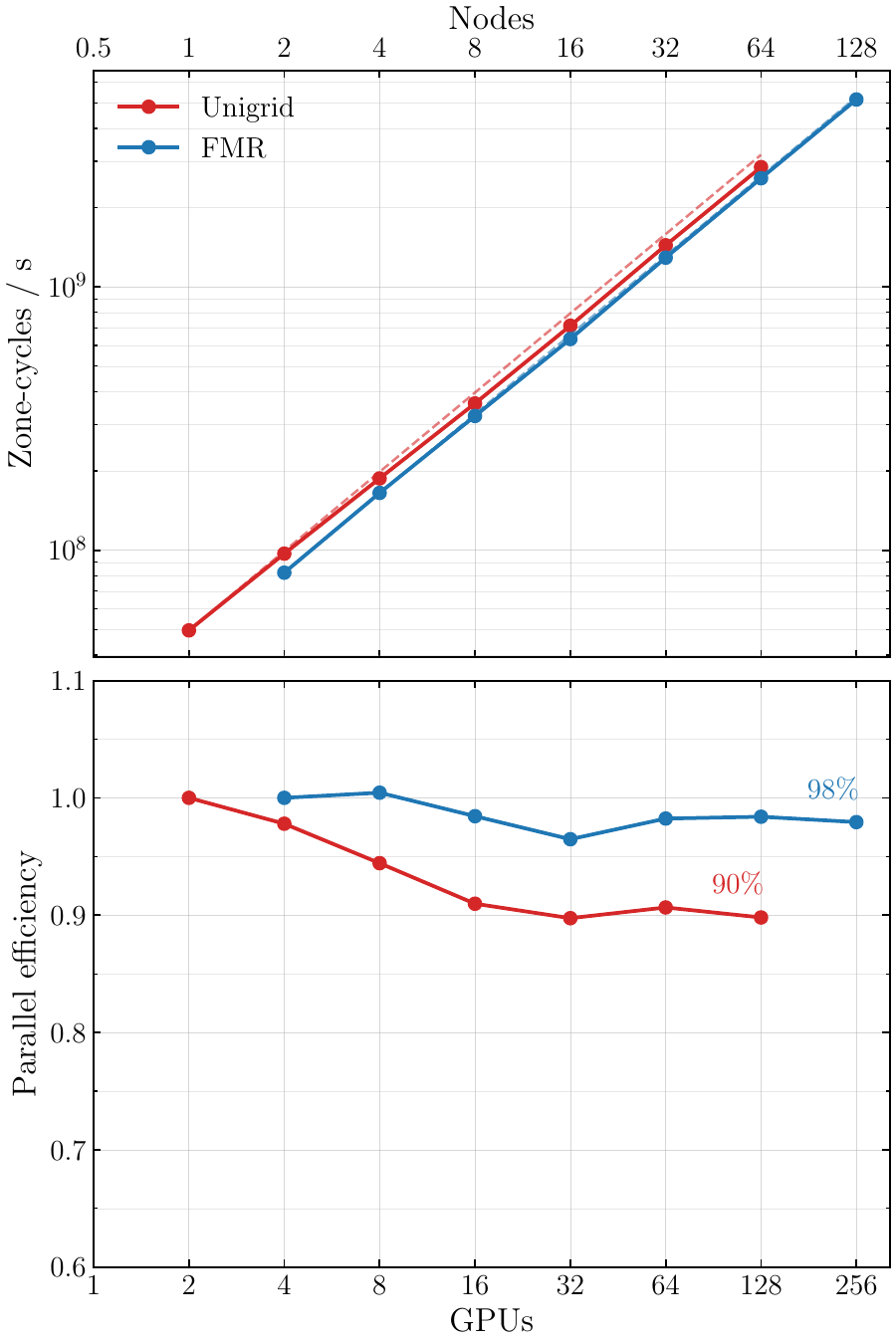}
\caption{Weak scaling of \GRACE\ on the Viper cluster for the uniform-grid (red) and FMR (blue) A0 TOV configurations. \textit{Top:} zone-cycles per second as a function of device (and node) count; \textit{bottom:} parallel efficiency, which remains ${\simeq}\,90\%$ for the uniform-grid setup (out to $128$ devices) and ${\simeq}\,98\%$ for the FMR setup (out to $256$ devices).}\label{fig:weak_scaling}
\end{figure}

As shown in Fig.~\ref{fig:weak_scaling}, \GRACE\ achieves an efficiency $\simeq 90\%$ for the unigrid case when going from $2$ to $128$ devices (red filled circles) and $\simeq 98\%$ for the FMR setup when going from $4$ to $256$ devices (blue filled circles). This indicates that the ghost-exchange task graph successfully hides the inter-device communication cost in the regime where the per-device load is large enough to saturate the kernel-launch and memory-access latencies.  We attribute the lower efficiency in the unigrid case to two factors. Firstly, the grid we employ does not fully saturate the device memory; secondly this simpler setup does not perform any prolongation or restriction, making the ghost-filling operation computationally much cheaper. Both of these aspects cause the MPI communication to have a higher relative weight on the total time taken to perform an iteration.

To conclude this section, and as an indication of the cost of running the code in production conditions, we will now describe the performance of the code in the realistic BNS simulation of Sec.~\ref{ssec:bns}. The LR, MR, and HR runs presented there were performed on $4$, $8$, and $24$ AMD MI300A Accelerated Processing Units (APUs), respectively, on the Viper cluster at the Max Planck Computing and Data Facility. The average sustained evolution rates were $\sim 6.4\times 10^2$, $\sim 3.3\times 10^2$, and $\sim 3.05\times 10^2\,M_\odot/\mathrm{h}$ of physical time per walltime hour throughout the inspiral. This implies that the three simulations took approximately $\sim 4.9$, $\sim 9.9$ and $\sim 10.8$ walltime hours to reach merger, respectively, corresponding to a computational cost of $\sim 20$, $\sim 79$ and $\sim 259$ ${\rm APU-h}$ for the resolution triplet. Since the merger time is similar across resolutions (see the discussion in Sec.~\ref{ssec:bns}) and the code is $3+1$ dimensional with the time-step scaling with the grid spacing, a rough scaling model would imply that the cost should grow with resolution as $h^{-4}$. The measured cost scales as $h^{-3.4}$ (LR--MR) and $h^{-4.1}$ (MR--HR), bracketing the simple $h^{-4}$ expectation. We stress that these are crude two-point estimates at differing APU counts, so they convolve algorithmic cost with parallel efficiency. Nonetheless, their consistency with $h^{-4}$ indicates no gross departure from ideal scaling under production conditions, complementing the controlled measurements of the rest of this section. The low-to-medium resolution ratio being lower than expected is likely an indication that the LR setup was slightly under-loaded on $4$ APUs.

\section{Conclusions}
\label{sec:conclusions}

We have introduced \texttt{GRACE}, a new GPU-accelerated GRMHD framework, developed from scratch and built exclusively on open-source libraries, including \texttt{Kokkos}~\cite{Trott2021,Trott2022_etal} for performance portability across CPU and GPU backends and \texttt{p4est}~\cite{Burstedde2011, carsten_burstedde_2024_10839051} for the management of adaptively refined grids. \texttt{GRACE} solves the equations of ideal GRMHD---with divergence-free magnetic fields maintained by constrained transport---self-consistently coupled to the Einstein equations in the Z4c formulation, on fixed or adaptively refined meshes. The code is publicly available, together with the companion Python library \texttt{GRACEpy} for post-processing and data analysis (see Sec.~\ref{sec:availability}).

We demonstrated the capabilities of \texttt{GRACE} through a sequence of tests spanning the range of regimes the code targets: from the GRMHD solver in flat spacetime (magnetized shock tubes and the magnetic rotor), through GRMHD on a fixed curved background (magnetized Bondi accretion onto a Schwarzschild black hole and neutron-star oscillation spectra in the Cowling approximation) and the vacuum Z4c sector (the ringdown of a perturbed spinning puncture, cross-checked against the \texttt{FIL} code, and binary black-hole inspiral--merger--ringdown waveforms), to the fully coupled system (neutron-star oscillation spectra in dynamical spacetimes).

We further validated the implementation on two binary neutron-star mergers: an equal-mass, unmagnetized system evolved with an ideal-gas EOS, and an unequal-mass ($q=0.8$), magnetized system evolved with the finite-temperature, tabulated SFHo EOS. For the latter we compared the inspiral dynamics against the \texttt{FIL} code and found broad agreement, with the residual phase difference consistent with the two codes' intrinsic discretization errors.

We also characterized the parallel performance of the code across multiple accelerator architectures. On the FMR benchmark of Sec.~\ref{ssec:single_device}, \texttt{GRACE} reaches single-device throughputs of $2.48\times 10^7$ and $1.05\times 10^7$ zone-cycles per second for fixed-spacetime GRMHD evolutions on an AMD MI300A APU and an NVIDIA A100 GPU, respectively, and $1.68\times 10^7$ and $6.7\times 10^6$ zone-cycles per second for the fully coupled GRMHD--Z4c system. On the Viper cluster it retains strong-scaling efficiencies of ${\sim}44\%$ on uniform grids over a sixteenfold increase in device count and ${\sim}54\%$ on FMR grids over an eightfold increase, and weak-scaling efficiencies of ${\simeq}90\%$ out to $128$ devices on uniform grids and ${\simeq}98\%$ out to $256$ devices on FMR grids.

Future releases will extend \texttt{GRACE} with radiative transport and microscopic treatments of weak interactions, as well as effective descriptions of dissipative effects beyond ideal MHD~\cite{Palenzuela_2013,Franceschetti_2025}.

\section{Acknowledgements}
\label{sec:acknowledgements}

CM would like to thank Sebastiano Bernuzzi, Boris Daszuta, Tim Dietrich, Yong Gao, Alan Tsz-Lok Lam, and Masaru Shibata for insightful discussions during the development of \texttt{GRACE}; Roberto De Pietri for insightful discussions and for sharing data used in cross-code comparisons; Kenta Kiuchi and Alexis Reboul-Salze for their help with the MHD scheme; Harald Pfeiffer for his advice on gravitational-wave data analysis; and Ming-Zhe Han for his advice on the optimization of the metric evolution equations on GPU; as well as Martin Bernreuther and the HLRS team for support during the early phases of development. KT would like to thank Martin Ohlerich for his assistance regarding the software and compilation on SuperMUC-NG2. CE acknowledges support by the DFG through the CRC-TR 211 ``Strong-interaction matter under extreme conditions''---project number 315477589---TRR 211. MC, CM, and LR acknowledge support from the ERC Advanced Grant ``JETSET: Launching, propagation and emission of relativistic jets from binary mergers and across mass scales'' (grant no. 884631). LR also acknowledges the Walter Greiner Gesellschaft zur F\"orderung der physikalischen Grundlagenforschung e.V. through the Carl W. Fueck Laureatus Chair. 
ERM acknowledges support by the U.S. National Science Foundation under Grants No. PHY-2541792 and PHY-2309210. ERM is also supported by a Research Fellowship from the Sloan Foundation, and a William H. Hurt Scholarship at Caltech.
The simulations and tests presented in this paper were performed on the GPU partitions of the Viper and Raven clusters at the Max Planck Computing and Data Facility. Part of the code development was carried out on the GPU partition of the Goethe-NHR cluster, as well as on Hawk-AI and Hunter at HLRS, within the ``GRACE'' Test and Development allocation (Project ID: 44305). Comparison runs with \texttt{FIL} using CPU resources were performed on MareNostrum5 GPP of the EuroHPC allocation EHPC-REG-2025R02-175 ``Pushing the edge in compact binary mergers: Inspiral, Instabilities, and Radiation-Driven Outflows''. \texttt{ChatGPT~5.2} was used to assist with the design of plotting scripts and to provide enhanced spelling and grammar checking. CM used \texttt{Claude Opus 4.7} and \texttt{4.8} (1M-context) in the final stages of development and validation of \texttt{GRACE} and in the writing of this manuscript.

\section{Data and Code Availability}
\label{sec:availability}

The \texttt{GRACE}~code presented in this paper is open source and can be downloaded at \href{https://github.com/GRACE-astro/grace}{https://github.com/GRACE-astro/grace}. An accompanying set of Python scripts for data analysis and post-processing, collected in the environment \texttt{GRACEpy}, is available at \href{https://github.com/GRACE-astro/GRACEpy}{https://github.com/GRACE-astro/GRACEpy}.

\bibliography{main.bib}

\end{document}